\documentclass[useAMS,usenatbib, times]{mn2e}

\def\hsp{\!\!\!\!}
\def\const{0.006}
\def\dr{{\rm{d}}}
\def\eabs{\vert E\vert}
\def\ecc{\epsilon}
\def\gyr{{\rm Gyr}}
\def\henon{H\'enon}
\def\kms{{\rm km\,s}^{-1}}
\def\kpc{{\rm kpc}}
\def\mdot{\dot M}
\def\mdotabs{\vert\dot M\vert}
\def\mdotx{\hat{M}_x}
\def\mmax{m_{\rm max}}
\def\mmean{\bar m}
\def\mmin{m_{\rm min}}
\def\mref{M_{\rm 1}}
\def\msun{{\rm M}_{\odot}}
\def\msunpc{\msun{\rm pc}^{-3}}
\def\myr{\rm Myr}
\def\ndot{\dot N}
\def\nref{N_1}
\def\pc{{\rm pc}}
\def\rg{R_{\rm G}}
\def\rh{r_{\rm h}}
\def\rhdot{\dot r_{\rm h}}
\def\rhoh{\rho_{\rm h}}
\def\rhoj{\rho_{\rm J}}
\def\rhrjmax{\left(\frac{\rh}{\rj}\right)_1}
\def\rhrjmaxtext{(\rh/\rj)_1}
\def\rhrjmaxtexts{[\rh/\rj]_1}

\def\rhmax{\hat\rh}
\def\rj{r_{\rm J}}
\def\rp{R_{\rm P}}
\def\tcr{\tau_{\rm cr}}
\def\tcrdot{\dot\tau_{\rm cr}}
\def\tcrinit{\tau_{\rm cr0}}
\def\tcrj{\tau^{\rm J}_{\rm cr}}
\def\tcrmax{\tau_{\rm cr1}}
\def\tev{\tau_{\rm ev}}
\def\tevn{\tau_{\rm ev0}}
\def\thubble{t_{\rm H}}
\def\tml{\tau_{\rm esc}}
\def\trh{\tau_{\rm rh}}

\def\trhmax{{\hat\tau}_{\rm rh}}
\def\vg{V_{\rm c}}
\def\vs{{\sl vs.}}

\long\def\symbolfootnote[#1]#2{\begingroup%
\def\thefootnote{\fnsymbol{footnote}}\footnote[#1]{#2}\endgroup} 
\def\aj{AJ}%          % Astronomical Journal
\def\araa{ARA\&A}%          % Annual Review of Astron and Astrophys
\def\apj{ApJ}%          % Astrophysical Journal
\def\apjl{ApJ}%          % Astrophysical Journal, Letters
\def\apjs{ApJS}%          % Astrophysical Journal, Supplement
\def\aap{A\&A}%          % Astronomy and Astrophysics
\def\mnras{MNRAS}%          % Monthly Notices of the RAS
\usepackage{amssymb,latexsym,graphicx,natbib,eufrak,times}

\title[The life cycle of star clusters]
  {The life cycle of star clusters in a tidal field}
\author[Gieles, Heggie \& Zhao]
  {Mark~Gieles$^1$, Douglas~C. Heggie$^2$, HongSheng Zhao$^3$\\
$^1$ Institute of Astronomy, University of Cambridge, Madingley Road, Cambridge, CB3 0HA, UK\\
 $^2$ School of Mathematics and Maxwell Institute for Mathematical Sciences, University of Edinburgh, KingÕs Buildings, Edinburgh EH9 3JZ\\
 $^3$Scottish University Physics Alliance, University of St. Andrews, KY16 9SS, UK
}
\date{Accepted 2011 January 9.  Received 2011 January 9; in original form 2010 November 22}

\def\LaTeX{L\kern-.36em\raise.3ex\hbox{a}\kern-.15em
    T\kern-.1667em\lower.7ex\hbox{E}\kern-.125emX}

\begin{document}         
\maketitle
\begin{abstract}
The evolution of globular clusters due to 2-body relaxation results in
an outward flow of energy and at some stage all clusters need a
central energy source to sustain their evolution. \henon\ provided the
insight that we do not need to know the details of the energy
production in order to understand the relaxation-driven evolution of
the cluster, at least outside the core.  He provided two self-similar
solutions for the evolution of clusters based on the view that the
cluster as a whole determines the amount of energy that is produced in
the core: steady expansion for isolated clusters, and homologous
contraction for clusters evaporating in a tidal field. The amount of
expansion or evaporation per relaxation time-scale is set by the
instantaneous radius or number of stars, respectively.  We combine
these two approximate models and propose a pair of Unified Equations
of Evolution (UEE) for the life cycle of initially compact clusters in
a tidal field.  The half-mass radius increases during the first part
(roughly half) of the evolution, and decreases in the second half;
while the escape rate approaches a constant value set by the tidal
field.  We refer to these phases as `expansion dominated' and
`evaporation dominated'.  These simple analytical solutions of the UEE
immediately allow us to construct evolutionary tracks and isochrones
in terms of cluster half-mass density, cluster mass and
galacto-centric radius. From a comparison to the Milky Way globular
clusters we find that roughly one-third of them are in the second,
evaporation-dominated phase and for these clusters the density inside
the half-mass radius varies with the galactocentric distance $\rg$ as
$\rhoh\propto\rg^{-2}$. The remaining two-thirds are still in the
first, expansion-dominated phase and their isochrones follow the
environment-independent scaling $\rhoh\propto M^2$, where $M$ is the
cluster mass; that is, a constant relaxation time-scale. We find
substantial agreement between Milky Way globular cluster parameters
and the isochrones, which suggests that there is, as H\'enon
suggested, a balance between the flow of energy and the central energy
production for almost all globular clusters.
\end{abstract}

%%%%%%%%%%%%%%%%%%%%%%%%%%%%%%%%%%%%%%%%%%%%
\section{Introduction}\label{sec:intro}
\begin{figure}
\begin{center}
\includegraphics[width=8.cm]{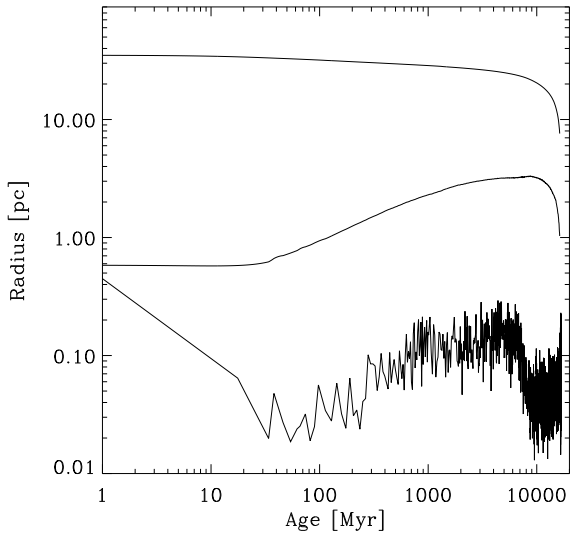} 
\end{center}
     \caption{The evolution of the tidal, half-mass and core-radii
       (from top to bottom) in the evolution of a Monte Carlo model of
       the globular cluster M4 \citep[from][]{2008MNRAS.389.1858H}.  }
   \label{fig:mc-model}
\end{figure}

Globular clusters strive their entire life for thermal equilibrium,
but their negative heat capacity prevents them from ever reaching
this. As a result there is a continuous flow of energy that is
conducted outwards by relaxation. Any equations attempting to capture
cluster evolution could be very complex and non-linear with many
variables to keep track of since the dynamical evolution of star
clusters is the result of several processes, including two-body
relaxation, interactions with binary stars, escape across the tidal
boundary, and the internal evolution and mass-loss of single and
binary stars. Fig.\ref{fig:mc-model} illustrates typical results, in
terms of the evolution of the core radius, the half-mass radius and
the tidal radius.  Within the first few times $10$\,Myr there is a
rapid phase of mass segregation, leading to contraction of the core.
(Note the logarithmic time scale in the figure.)  At the end of this
phase the half-mass radius starts to increase, and the expansion
continues, but decelerates, over the next 8\,Gyr approximately.  At
the same time the cluster is losing mass, both by the escape of stars
and by the effects of stellar evolution, and this causes a decrease in
the tidal radius.  For the remaining evolution (up to 12\,Gyr) the
half-mass radius also contracts.  Except for the initial phase of mass
segregation, the core radius remains small compared with the half-mass
radius; it also begins its contraction at about 6\,Gyr, i.e.  earlier
than the half-mass radius, and subsequently drops to and fluctuates
around a lower value.

Our theoretical aim in this paper is to provide a physically motivated
and simple prescription for the behaviour of the half-mass and tidal
radii in the period following the end of the phase of mass
segregation, i.e. all except about the first 1\% of the evolution of
the cluster.  With this we are able to construct evolutionary tracks
and isochrones for clusters evolving in a tidal field and provide a
theoretical framework for empirically established correlations between
structural parameters and their environment as found for Milky Way
globular clusters
\citep[e.g.][]{1995ApJ...438L..29D,2000ApJ...539..618M} and for
extra-galactic globular cluster systems
\citep[e.g.][]{2005ApJ...634.1002J,2007AJ....133.2764B,2008MNRAS.384..563M,2010MNRAS.401.1965H}. We
do not aim to explain the shape and dependence on environment of the
globular cluster mass function at this stage, however.  The other
principal exclusion is that we do not attempt to model the evolution
of the core.  The fact that this programme will succeed is due to a
number of important discoveries by \henon.

The first of these discoveries \citep{1961AnAp...24..369H} was a
simple model of an idealised star cluster evolving in a tidal field.
The main idealisations were that all stars have the same mass and do
not evolve internally, the tidal field is modelled as a cutoff at the
tidal radius, the system is non-rotating and spherically symmetric,
there is no significant binary population, the velocity distribution
of the stars at a point is isotropic, and gravitational encounters are
modelled as a Fokker-Planck equation.  Then \henon\ showed that the
system could evolve at constant mean density, with stars escaping
across the tidal boundary, and the structure of the cluster at any
time is simply a scaled version of the structure at any other time,
i.e. the evolution is `self-similar'.  In particular, this means that
the half-mass radius is a constant fraction of the tidal radius.  We
shall adapt this model to describe the evolution at late times in
models such as that shown in Fig.\ref{fig:mc-model}.

\henon's next advance in this field \citep{1965AnAp...28...62H} was
the discovery of a similar model for an isolated system, i.e. one with
no tide. It is again self-similar, but the total mass remains
constant, and the half-mass radius increases.  This we shall adapt to
describe the evolution of models such as that shown in
Fig.\ref{fig:mc-model}, from the end of the phase of mass segregation
until the point where the tidal radius begins to have a significant
effect on the expansion of the half-mass radius.  A further part of
our adaptation will be to construct a way of bridging the transition
between the two models, and to do so in a physically well motivated
way.

The third discovery of \henon\ explains why this programme can work
even without detailed modelling of the core.  In both of \henon's
models the energy of the system increases, and \henon\ himself
considered that the formation and hardening of binaries in the core
was a plausible source of this energy\footnote{In fact
  \citet{1961AnAp...24..369H} introduces the concept of energy
  generation by the core in the other way around: the core absorbs
  negative energy.}.  It might have been thought, therefore, that the
evolution of the system would depend on details of this source of
energy and the radius and density of the core.  But a correct
understanding of the connection between the core and the evolution of
the system as a whole came from the insight of
\citet{1975IAUS...69..133H}.  He realised that the {\it rate of flow
  of energy is controlled by the system as a whole, and not by the
  core.}  That is, it is not the cluster that responds to whatever
happens in the core, but the other way around.

In \henon's picture the mechanism of energy generation is
self-regulatory, as is the case in stellar interiors, where the
luminosity of the star is determined by how much radiation can be
transported through the envelope. The nuclear fusion rate in the core
of the star adjusts to it. The application of this idea to stellar
dynamics was a breakthrough allowing modellers to overcome the core
collapse phase. In the Monte Carlo models of
\citet{1975IAUS...69..133H} the energy source is entirely artificial
and unspecified.  Later, it was found from $N$-body simulations of the
long term (post-collapse) evolution of equal mass clusters that hard
binaries act as the energy source
\citep{1994MNRAS.268..257G,2002MNRAS.336.1069B}.  In more realistic
$N$-body simulations of star clusters these binaries are usually
considered to be primordial, but other mechanisms of energy generation
have been modelled, including the action of a central
intermediate-mass black hole \citep{2004ApJ...613.1143B} and, less
controversially, mass-loss from stellar evolution.
\citet{2010MNRAS.408L..16G} showed that  mass loss as a consequence
of stellar evolution also leads to a self-regulatory energy
production, which works together with hard binaries in driving the
long term evolution.  Indeed, it may even dominate, as in the specific
example of 47 Tuc \citep{2011MNRAS.410.2698G}, a high-concentration
cluster in which the evolution of the core- and half-mass radii
appears to be little affected by the primordial binary population.

These self-regulatory mechanisms of energy generation take time to
establish the balance between the energy generated in the core and the
energy requirements of the overall evolution of the cluster.  In
Fig.\ref{fig:mc-model} this balance is reached at about the end of the
phase of mass segregation.  Mass segregation ceases close to the time
when the core radius reaches a minimum for the first time.  This was
found from numerical simulations of clusters with a moderately wide
mass spectrum \citep[][]{1996MNRAS.279.1037G} and for clusters with a
full mass spectrum \citep{2007MNRAS.378L..29P}.  The reason for this
is not that equipartition has been reached, due to Spitzer's
instability \citep{1969ApJ...158L.139S}.  The origin of this quasi
steady-state distribution of stars of different masses is not well
understood theoretically, but it is taken as a given here.  We refer
to the subsequent evolution as `balanced'.  In much research on
cluster dynamics this kind of evolution is usually associated with
`post-collapse' evolution, but this term is ambiguous; in
Fig.\ref{fig:mc-model}, for instance, the phrase `post-collapse' might
be used for the entire evolution after the end of mass segregation,
but others might apply it to the evolution following the late decrease
in the core radius in the last few Gyr.  For this reason we prefer the
term `balanced evolution'.

\citet{2010MNRAS.408L..16G} showed that the evolution of most globular
clusters is balanced.  Their result was based, however, on a
comparison of the behaviour of clusters expanding in isolation to the
structural parameters of relatively massive globular clusters
($\gtrsim$few$\times10^4\,\msun$) and ultra-compact dwarf galaxies.
It was assumed implicitly that these are little affected by the tide.
In the present study we add the effect of a tidal field, which slows
down the expansion of all clusters at some point in their evolution.

In Section~\ref{sec:cycle} we discuss how the behaviour of the
\henon\ models in isolation and in tidal fields can be related to the
flow of energy, and we show how the two extremes, expansion without
mass loss and homologous contraction in a tidal field, may be
unified. Isochrones in physical units at an age of a Hubble time are
extracted and are compared to the data of Milky Way globular clusters
in Section~\ref{sec:obs}. A summary and discussion are presented in
Section~\ref{sec:discussion}. The specific symbols used in this study
are summarised in Table~\ref{tab:acronyms} and details of the
derivations are presented in Appendices.

\begin{table*}
\caption{Overview of the specific symbols used in this study. The
  symbols with the subscript $1$ are used to refer to the maximum
  value the variable can have.}
\label{tab:acronyms}
\begin{tabular}{lll}
\hline
Symbol         & Definition                        & Description\\\hline
$\rhoh$        &$3M/(8\pi\rh^3)$                   & Mean density within the half-mass radius\\
$\rhoj$        &$3M/(4\pi\rj^3)$                   & Mean density within the Jacobi radius\\
$\rhrjmaxtext$ &                                   & Maximum ratio of the half-mass radius over the Jacobi radius\\
$\tcr$         & $(G\rhoh)^{-1/2}$                  & A measure of the crossing time at the half-mass radius\\
$\tcrj$        & $(G\rhoj)^{-1/2}$                  & Crossing time at the Jacobi radius\\
$\tcrmax$      & $\sqrt{2}\rhrjmaxtext^{3/2}\tcrj$  & Crossing time at the half-mass radius for a homologous cluster, equation~(\ref{eq:tcrmax})\\
$\trh$         & $CN\tcr$                    & Approximate half-mass relaxation time,  with $C\simeq\const$ (equations~\ref{eq:trhapprox}\,\&\,\ref{eq:ctrh}) \\
$\tevn$        & see equation~(\ref{eq:tev})       & Total life-time of a cluster of $N_0$ stars\\
$\tev$         &  see equation~(\ref{eq:tev})      & Remaining life-time, or `life-expectancy', of a clusters of $N$ stars \\
$\zeta$        & $(\dot{E}/\eabs)\trh$             & Fraction of the total energy conducted per $\trh$\\
$\chi$         & $(\tcrdot/\tcr)\trh$              & Dimensionless expansion rate\\
$\xi$          & $-(\ndot/N)\trh$                  & Dimensionless escape rate\\
\hline
\end{tabular}
\end{table*}

%%%%%%%%%%%%%%%%%%%%%%%%%%%%%%%%%%%%%%%%%%%%
\section{A unified model of expansion and evaporation}
\label{sec:cycle}

 %________________________________________________
\subsection{The flux of energy}
\label{ssec:eflux}

Our attempt to unify the evolution of mass and radius of the two
models of \henon\ begins with one of the physical properties which the
two models have in common, i.e. a flux of energy at the half-mass
radius which is fed by an energy source in the core.  We constrain
ourselves to the energy flow at this radius, because we can then
construct a relatively simple set of relations for the behaviour of
the bulk properties of the cluster.  A full unification of the two
models of \henon\ requires numerically solving the Fokker-Planck
equations \citep[e.g.][and follow-up
  studies]{1979ApJ...234.1036C}. This would be needed to get the
distribution function of energies and the density profile as a
function of time, which is not the aim of this paper.  For the present
we adopt \henon's idealisation of systems in which all stars have the
same mass $m$.  Then we estimate the energy as usual by

\begin{equation}
E =  -\alpha\frac{GN^2m^2}{\rh},\label{eq:E}
\label{eq:e}
\end{equation}
where $N$ is the number of stars, $\rh$ is the half-mass radius, and
$\alpha$ is a `form factor' for which the value 0.2 is often taken
\citep[p.12]{1987degc.book.....S}.

In \henon's isolated model \citep{1965AnAp...28...62H}, all quantities
on the right-hand side of equation~(\ref{eq:E}) are constant except
$\rh$, and it follows from \henon's equations~(3), (6) and (7) that

\begin{equation}
  \frac{\dot E}{\vert E\vert} = \frac{\dot\rh}{\rh} \simeq \frac{0.0926}{\trh},
\label{eq:flux-isolated}
\end{equation}
except that one also requires the value for the dimensionless
half-mass radius $\mathbf{R}$, which is given in \henon's paper on
p.64.  Also, we have expressed the result in terms of the conventional
half-mass relaxation time \citep{1987degc.book.....S}, i.e.

\begin{equation}
\trh=0.138\frac{N^{1/2}\rh^{3/2}}{\sqrt{\bar{m}G}\ln\Lambda},
\label{eq:trh_spitzer_hart} \label{eq:trh}
\end{equation}
where $\bar{m}$ is the mean stellar mass (i.e. $m$ in this case) and
$\ln\Lambda$ is the Coulomb logarithm, which we have equated with the
factor $\mbox{Log~}n$ in \henon's equation~(7).

For \henon's tidally limited model \citep{1961AnAp...24..369H}, both
$N$ and $\rh$ vary because the cluster loses stars at a constant
density, and a similar calculation leads to the result that

\begin{equation}
  \frac{\dot E}{\vert E\vert} = -2\frac{\dot N}{N} +
  \frac{\dot\rh}{\rh} \simeq
  \frac{0.0743}{\trh}.  
\label{eq:flux-tidal-henon}
\end{equation}

What is noticeable about equations~(\ref{eq:flux-isolated}) and
(\ref{eq:flux-tidal-henon}) is how similar the final numerical
coefficients are.  Indeed, in constructing a unified approximate model
which includes the transition from nearly isolated evolution to tidally
limited evolution, we shall make the {\it assumption} that the
numerical coefficient is constant, i.e. that

\begin{equation}
  \frac{\dot E}{\vert E\vert} = \frac{\zeta}{\trh},
\end{equation}
where $\zeta \simeq 0.08$ in the case of equal masses.  The accuracy
of this approximation is comparable with the common approximation of
treating $\alpha$ in equation~(\ref{eq:E}) as constant.

Before proceeding further, we shall change one of the variables in
which the total energy $E$ is expressed, because this will facilitate
one particular step in the further development of our model.  Instead
of using the half-mass radius, $\rh$, we shall use the (half-mass)
crossing time of the system, defined here as
\begin{equation}
\tcr\equiv\left(G\rhoh\right)^{-1/2},
\label{eq:tcr}
\end{equation}
with $\rhoh\equiv3M/(8\pi\rh^3)$, the cluster density within the
half-mass radius, and $M\equiv m N$ the total cluster mass.  (Note
that this differs from the conventional crossing time by a numerical
factor; see also Table \ref{tab:acronyms}.)  Putting this together
with our assumption about the energy flux, we replace
equations~(\ref{eq:flux-isolated}) and (\ref{eq:flux-tidal-henon}) by
\footnote{For clusters with a stellar mass spectrum there will be an
  additional term $-(5/3)(\dot{\mmean}/\mmean)$ on the right-hand
  side.}
\begin{equation}
  \frac{\dot E}{\vert E\vert}  =  \frac{\zeta}{\trh}= -\frac{5}{3}\frac{\dot N}{N} +
  \frac{2}{3}\frac{\dot\tcr}{\tcr}.  \label{eq:flux-tidal}
\label{eq:edottot}
\end{equation}
 The constants become obvious when we look at the expression for
  the total energy (equation~\ref{eq:e}) in terms of $N$ and $\tcr$
  (equation~\ref{eq:tcr}): $\eabs\propto (mN)^{5/3}/\tcr^{2/3}$. In the
next subsection the relative contributions of expansion and
evaporation to the energy flow are specified.

 %________________________________________________
\subsection{The relative importance of expansion and evaporation}
\label{ssec:relative}
Now we have a functional form that relates evaporation and expansion
to the flow of energy (equation~\ref{eq:edottot}) we need to specify
what the relative importance is of these two processes. The equations
that follow (equaions~\ref{eq:ndot}--\ref{eq:chi}) form the base for our
analysis. We introduce two new coefficients, $\xi$ and $\chi$, to
quantify the (dimensionless) evaporation rate and the (dimensionless)
expansion rate, respectively.

\begin{equation}
\xi\equiv-\frac{\ndot\trh}{N},
\label{eq:ndot}
\end{equation}
and
\begin{equation}
\chi\equiv\frac{\tcrdot\trh}{\tcr},
\label{eq:tcrdot}
\end{equation} 
so that (see equation~\ref{eq:edottot})
\begin{equation}
\zeta=\frac{2}{3}\chi+\frac{5}{3}\xi. 
\label{eq:zeta}
\end{equation}
For \henon's isolated model, $\xi = 0$, while $\chi = 0$ for the
tidally bound model.

To get a full description of the behaviour of the mass and half-mass
radius of the cluster we need to relate one of the coefficients $\xi$
or $\chi$ to $\tcr$ and $N$. Here we focus on the dimensionless
escape-rate $\xi$.  By estimating the fraction of stars above the
escape velocity for tidally limited clusters with different ratios
$\rh/\rj$, where $\rj$ is the Jacobi (tidal) radius of the cluster,
\citet{2002IAUS..207..584L} and \citet{2008MNRAS.389L..28G} have shown
that $\ln\xi\propto\rh/\rj$, approximately.  In principle we could
proceed with this exponential function for $\xi$, but a more
convenient approximation will make what is to follow much easier.  In
fact, \citet{2008MNRAS.389L..28G} show that
$\xi\propto(\rh/\rj)^{3/2}$ provides a satisfactory approximation to
the exponential function (in the relevant range of $\rh/\rj$). This
relation can be expressed equivalently as $\xi\propto\tcr$. As a
cluster expands, $\tcr$ increases, until the point where the cluster
evolution becomes tidally limited, i.e. the point where its evolution
begins to be analogous to the model of \citet{1961AnAp...24..369H}.
Let $\tcrmax$ denote the value of $\tcr$ during this part of the
evolution.  Since $\tcr$ is now constant, $\chi = 0$ (by
equation~\ref{eq:tcrdot}), and so equation~(\ref{eq:zeta}) implies
that $\xi = (3/5)\zeta$ when $\tcr = \tcrmax$.  Putting these results
together with equation~(\ref{eq:zeta}), we arrive at a pair of Unified
Equations of Evolution (UEE)
\begin{eqnarray}
\xi&\equiv&-\frac{\ndot\trh}{N}=\frac{3}{5}\zeta\frac{\tcr}{\tcrmax},\label{eq:xi}\\
\chi&\equiv&\frac{\tcrdot\trh}{\tcr}=\frac{3}{2}\zeta\left(1-\frac{\tcr}{\tcrmax}\right),\label{eq:chi}
\end{eqnarray}
which we adopt henceforth. 
These two expressions for $\xi$ and $\chi$ capture the behaviour of
clusters for all values $0\le\tcr\le\tcrmax$. For $\tcr\ll\tcrmax$
they give the desired self-similar expansion ($\chi\simeq(3/2)\zeta =
$\,constant and $\xi\simeq0$) and the growth of $\tcr$ stops when
$\tcr=\tcrmax$ ($\chi=0$ and $\xi=(3/5)\zeta$).  In this regime we
assume that the radial scale of the cluster is set by the tidal
radius, and so the ratio of the half-mass to tidal radii takes a
constant value, which we denote by $\rhrjmaxtext$.  In his equal-mass
homological model \citet{1961AnAp...24..369H} found the value
$\rhrjmaxtext \simeq 0.145$.

\begin{figure}
 \includegraphics[width=8.cm]{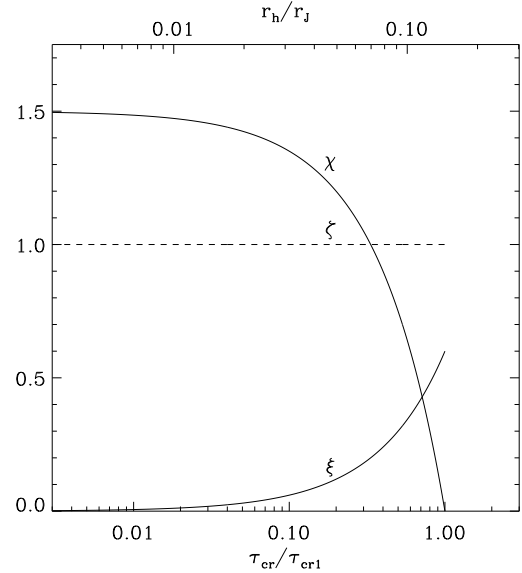}
     \caption{The dependence of the dimensionless escape rate $\xi$
       and expansion rate $\chi$ on $\tcr/\tcrmax$
       (equations~\ref{eq:xi} and \ref{eq:chi}, respectively). The
       values of $\chi$ and $\xi$ are given relative to the efficiency
       of energy conduction, $\zeta$. The top axis labels
       corresponding values of $\rh/\rj$ using
       equation~(\ref{eq:translate}) and $\rhrjmaxtext=0.145$.}
   \label{fig:xi}
\end{figure}

In Fig.~\ref{fig:xi} we show the behaviour of $\xi$ and $\chi$ in
units of $\zeta$ as a function of $\tcr/\tcrmax$. The top axis shows
how $\rh/\rj$ relates to $\tcr/\tcrmax$, which will be explained in
more detail in the next subsection (equation~\ref{eq:translate}).

 %________________________________________________
\subsection{Time scales}

Before proceeding to solve the equations of our model for cluster
evolution, it is convenient to collect one or two straightforward
formulae for future reference.  First, we often approximate the
  half-mass relaxation time (equation~\ref{eq:trh}) by

  \begin{equation}
    \trh \simeq CN\tcr\label{eq:trhapprox},
  \end{equation}
 where  
  \begin{eqnarray}
    C &= &\left(\frac{3}{8\pi}\right)^{1/2}\frac{0.138}{\ln\Lambda},\nonumber\\
        &\simeq& \const.
  \label{eq:ctrh}
  \end{eqnarray}
The approximation in the last step 
 is approximately correct if we take $\Lambda = 0.02N$ \citep{1996MNRAS.279.1037G} with
$N=10^5$. 
We turn next to the crossing time $\tcr$, which is
defined in terms of the mean density inside the half-mass radius in
Table \ref{tab:acronyms}.  Since $\tcr\propto (\rh^3/M)^{1/2}$ and
$M\propto \rj^{3}$, it follows that $\tcr$ is related to
$\rh/\rj$ and its value in the evaporation-dominated regime as
  \begin{equation}
\frac{\tcr}{\tcrmax} =
  \displaystyle{\left(\frac{\rh/\rj}{\rhrjmaxtexts}\right)^{3/2}}.
\label{eq:translate}
  \end{equation}
At this point it is also convenient to give an explicit formula for
$\tcrmax$, which is 
\begin{equation}
  \tcrmax = \sqrt{\frac{8\pi\rj^3}{3GM}}\rhrjmax^{3/2}.
\label{eq:tcrmax}
\end{equation}

Table \ref{tab:acronyms} also introduces the `crossing time at the
Jacobi radius', which may be expressed as
  \begin{equation}
  \tcrj =\sqrt{\frac{4\pi\rj^3}{3GM}}.
  \label{eq:tcrj}    
  \end{equation}

Because we now have constrained the functional forms of $\chi$ and
$\xi$ and we have defined the time-scales involved we can solve for
the evolution of $\tcr(N)$ and the time-dependence of both $N$ and
$\tcr$, and we proceed to do this in the next subsection.
 
 %________________________________________________
\subsection{Motion in the $\tcr-N$ plane}
\label{subsec:tcr_n}
\begin{figure}
 \includegraphics[width=8.cm]{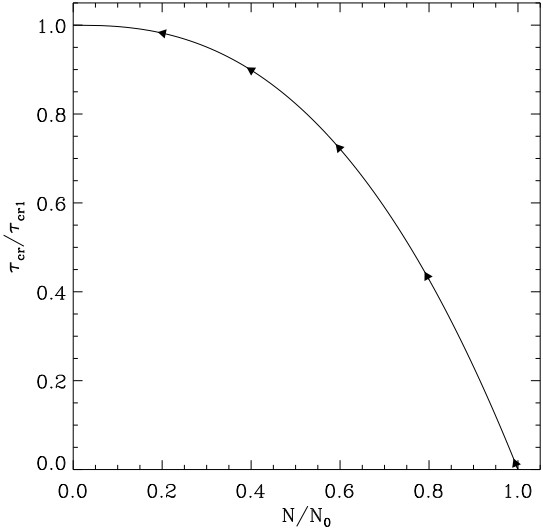} 
     \caption{Approximate relation between $\tcr$ and $N$, both
     normalised to their maximum values (equation~\ref{eq:tcr_n}).
     {Evolution is from right to left.}}
   \label{fig:tcr_n}
\end{figure}

The evolution of $N$ and $\tcr$ with respect to one another can be
found by combining the descriptions of $\ndot$ (equation~\ref{eq:xi})
and $\tcrdot$ (equation~\ref{eq:chi}) as derived in
Section~\ref{ssec:relative}.  This gives us the amazingly simple
differential equation
\begin{equation}
\frac{\dr N}{\dr\tcr}=\frac{2}{5}\frac{N}{\tcr-\tcrmax}.
\end{equation}
After separating the variables and integrating we find
\begin{equation}
\tcr=\tcrmax\left(1-\left[\frac{N}{N_0}\right]^{5/2}\right),
\label{eq:tcr_n}
\end{equation}
where $N_0$ is a constant.  Strictly, it is the value of $N$
corresponding to $\tcr = 0$.  Our model, however, is intended to
describe the phase of balanced evolution, which begins when the
cluster is compact, and $\tcr$ is small but non-zero.  But it is easy
to see from equation~(\ref{eq:tcr_n}) that, when $\tcr$ is small, $N_0
- N \simeq (2/5)(\tcr/\tcrmax)N_0$.  Therefore the value of $N$ at the
start of balanced evolution differs little from $N_0$ provided that
the value of $\tcr$ then is much smaller than
$\tcrmax$. Fig.~\ref{fig:tcr_n} displays the relation between $\tcr$
and $N$. The arrows indicate the direction of the motion along the
track. The arrows are placed at constant intervals of time and this
time dependence will be discussed in
Section~\ref{ssec:time-dependence}.

%___________________________________________________________________
\subsection{Time dependence}
\label{ssec:time-dependence}
Here we complete the solution of equations~(\ref{eq:ndot}) and
(\ref{eq:tcrdot}). The most compact expression of our results will
involve the total evolution in time, from the moment when $\tcr = 0$
until the dissolution of the cluster when $N = 0$.  Realistic
applications of our model, however, require further consideration of
this time scale, and so we take the further development of the model
in two stages.

\subsubsection{In units of the evaporation time}
\label{ssec:tev}
The time dependence of $N$ and $\tcr$ can be solved by substituting
the expression for $\trh$ (equation~\ref{eq:trhapprox}) in the
expression for $\ndot$ (equation~\ref{eq:xi}), such that we find
\begin{equation}
\ndot=-\frac{(3/5)\zeta}{C\tcrmax},
\label{eq:ndottcrj}
\end{equation}
which is constant for a given galactic environment.  Thus we get the
simple and familiar expression for the time evolution of $N$
\begin{eqnarray}
N(t)&=&N_0+\ndot t, \\
  &=&N_0\,\left(1-\frac{t}{\tevn}\right), 
\label{eq:nt}
\end{eqnarray}
 where $\tevn$ is the total life time, which is the time it takes to reduce the
number of stars from $N_0$ to zero. This can be expressed as
\begin{eqnarray}
\tevn&\equiv& -N_0/\ndot,\nonumber\\
&=&\frac{\trhmax}{(3/5)\zeta},
\label{eq:tev}
\end{eqnarray}
where $\trhmax$ is the relaxation time of a tidally-limited cluster
with $N_0$ stars; i.e. we substitute $N = N_0$ and $\tcr = \tcrmax$,
giving
\begin{equation}
  \trhmax = CN_0\tcrmax.
  \label{eq:trhmax}
\end{equation}

The time evolution of $\tcr$ is now retrieved from the expression for
$N(t)$ given above and the expression for $\tcr(N)$
(equation~\ref{eq:tcr_n}), i.e.
\begin{equation}
\tcr(t)=\tcrmax\left(1-\left[1-\frac{t}{\tevn}\right]^{5/2}\right).
\label{eq:tcrt}
\end{equation}
A more familiar quantity, however, is $\rh$, whose evolution in time
simply follows from the combination of equations~(\ref{eq:translate}),
(\ref{eq:nt}) and (\ref{eq:tcrt}), i.e.
\begin{equation}
\rh(t)=\rhmax\,\left(1-\frac{t}{\tevn}\right)^{1/3}\left(1-\left[1-\frac{t}{\tevn}\right]^{5/2}\right)^{2/3}
\end{equation}
with $\rhmax\equiv{\rj}_0\rhrjmaxtext$.  Here ${\rj}_0$ is the Jacobi
radius of a cluster with $N_0$ stars, and $\rhmax$ can be thought of
as the half-mass radius the cluster would have had {if} it still
contained $N_0$ stars and had $\tcr=\tcrmax$.

Finally, the time evolution of $\trh$ (equation~\ref{eq:trhapprox}) is
\begin{equation}
\trh(t)=\trhmax\left(1-\frac{t}{\tevn}\right)\left(1-\left[1-\frac{t}{\tevn}\right]^{5/2}\right),
\label{eq:trhevolution}
\end{equation}
 with $\trhmax$ given in equation~(\ref{eq:trhmax}).

\begin{figure}
 \includegraphics[width=8.cm]{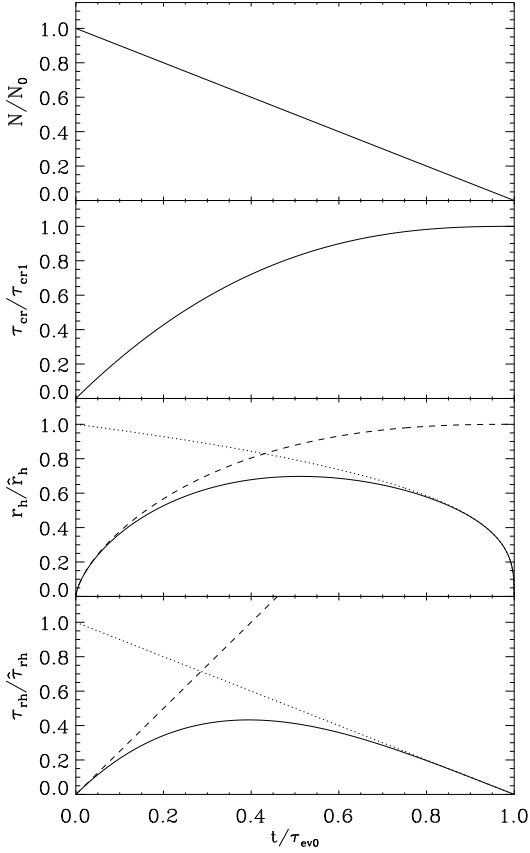} 
     \caption{Evolution of the parameters $N$, $\tcr$, $\rh$ and
       $\trh$ of clusters in a tidal field.  Time is normalised to the
       total evaporation time $\tevn$ and the normalisation
       constants on the $y$-scale are discussed in the text. The
       dashed lines in the bottom two panels show the extrapolation of
       the early evolution assuming that $N$ remains constant. The
       dotted lines show how $\rh$ and $\trh$ would have evolved if
       the evolution would have been homologous throughout:
       $\rh/\rj=\rhrjmaxtext=\,$constant.  }
   \label{fig:trh}
\end{figure}

The behaviour of $\tcr(t), N(t), \rh(t)$ and $\trh(t)$ is shown in
Fig.~\ref{fig:trh}.  The point where expansion and evaporation are
equally important, i.e. $\xi = \chi$, happens when
$t\simeq0.4\,\tevn$. (This is also the time at which the maximum value
of $\trh$ is reached, as $\trh \propto N\tcr$, if we neglect the
variation of the Coulomb logarithm.)  Therefore clusters are in an
expansion dominated phase for the first 40\% of their life, and in an
evaporation dominated phase in the following 60\%. In
Section~\ref{sec:obs} we will estimate the fraction of globular
clusters in the Milky Way that is in the expansion dominated phase.
Similarly, we can find the turning point in the evolution of $\rh$
from solving $\dot\rh=0$ which gives us
$t(\rhdot=0)\simeq0.5\tevn$. Therefore $\rh$ increases in the first
half of the cluster's life and decreases in the second half.

In the expansion phase, i.e. for $t\ll\tevn$, it follows from
equation~(\ref{eq:trhevolution}) that

\begin{equation}
  \trh(t) \simeq \frac{5}{2} \left(\frac{\trhmax}{\tevn}\right) t \simeq \frac{3}{2}\zeta t,    
  \label{eq:tsimtrh}
\end{equation}
where we have used equations~(\ref{eq:tev}) and (\ref{eq:trhmax}).
Hence we arrive at the interesting conclusion that the relaxation time
of all such clusters is the same at a given time. Furthermore, since
this relationship is represented by the dashed line in the lowest
panel of Fig.~\ref{fig:trh}, it is clear that this gives an upper
limit on the relaxation time at time $t$ for all clusters satisfying
the assumptions of our model.  We return to this point in
Section~\ref{ssec:isochrones}.

\subsubsection{In units of physical time}
\label{ssec:physical}
To describe the evolution in units of physical time we need to know
the value of $\tevn$, which by equations (\ref{eq:tcrmax}) and
(\ref{eq:tev}) depends essentially on $\zeta$, $\rhrjmaxtext$ and the
ratio $\rj^3/M$ (in addition to $N_0$).  The last of these three
quantities is set by the tidal field of the Galaxy, while the first
two parameters will be taken as constants, which we shall estimate by
comparison with numerical experiments of various kinds.

In isolated models we neglect escape, and then it is easy to show from
equations of Section~\ref{ssec:eflux} that
$(\rhdot/\rh)\trh\simeq(2/3)\chi\simeq\zeta$.
\citet{1998ApJ...495..786K} estimated the value of this quantity for
isolated two-component clusters with stars of masses $m_1$ and $m_2$,
and found, in our notation, that $\zeta$ is higher for larger ratios
$m_2/m_1$.  More quantitatively, \citet{2010MNRAS.408L..16G} use
direct $N$-body integrations to study the expansion of isolated
clusters with various mass functions and they find that the
dimensionless rate of expansion is approximately proportional to
$\mu^{1/2}$, where $\mu\equiv\mmax/\mmin$. For $\mu=10$, appropriate
for globular clusters, they find $\chi\simeq0.3$,
i.e. $\zeta\simeq0.2$.

The result that the efficiency of heat conduction (i.e. $\zeta$) is
larger in  multi-mass systems is confirmed by studies on the
escape rate of tidally-limited clusters.  Here we neglect $\chi$, and
the equations of Section~\ref{ssec:eflux} imply that the dimensionless
escape-rate is $(\ndot/N)\trh \simeq -\xi\simeq -(3/5)\zeta$.
\citet{1995ApJ...443..109L} consider various mass functions in their
Fokker-Planck simulations of dissolving clusters and study the effect
on the dimensionless escape rate.  Strictly, however, their results
relate to the rate of escape of mass, i.e. to $(\mdot/M)\trh$, but we
neglect the distinction here.  Then for $\mu=7$ they find that $\xi$,
and thus also $\zeta$, is a factor of $\sim$2 higher than for the
equal-mass case.  If we assume that $\xi\propto\mu^{1/2}$, then for
$\mu=10$, we again find $\zeta\simeq0.2$, a value which we adopt
henceforth. It is higher than for \henon's equal-mass models by a
factor of $\sim2.5$.

Rather than estimating $\rhrjmaxtext$ directly, we consider the ratio
$\zeta/\rhrjmaxtext^{3/2}$.  This can be done by computing the quantity
$\ndot\tcrj$ for it follows from equations~(\ref{eq:tcrmax}) and
(\ref{eq:ndottcrj}) that
\begin{eqnarray}
  \ndot\tcrj &=&-\frac{(3/5)\zeta}{\sqrt{2}C\rhrjmaxtext^{3/2}}.\\
            & \simeq&-\frac{71\,\zeta}{\rhrjmaxtext^{3/2}},
  \label{eq:tcrjndotn}
\end{eqnarray}
 where we used $C=\const$ in the last step (equation~\ref{eq:ctrh}).
This is the escape rate corrected for the tidal density, and should
thus be the same for clusters at different locations in a galaxy.  In
Table~\ref{tab:ndot} we summarise a number of results on this quantity
 from the literature, expressed in this way.

For \henon's equal-mass model ($\zeta\simeq0.0743$ and
$\rhrjmaxtext=0.145$, Sections~\ref{ssec:eflux} and
\ref{ssec:relative}) we find $\ndot\tcrj\simeq-95$. The Fokker-Planck
models of equal-mass clusters of \citet{1987ApJ...322..123L} start
relatively compact, and so nearly the entire evolution of their
clusters {is spent} in the post-collapse phase. Their results imply
that $\ndot\tcrj\simeq-76$. On the other hand their values for $\xi$
or $\zeta$ (i.e. the escape rate scaled by the half-mass relaxation
time, as in  equation~(\ref{eq:ndot}) are slightly larger than
\henon's value.  But from their Fig.~2 we see that in their models
$\rhrjmaxtext\simeq0.2$, which is slightly larger than {the value}
$\rhrjmaxtext=0.145$ in the model of \henon. This explains their
slightly lower value $\ndot\tcrj$ (equation~\ref{eq:tcrjndotn}). It
also illustrates that $\xi$ is not the only parameter that determines
the total life time in physical units.

Note that the results from studies based on $N$-body simulations in
Table \ref{tab:ndot} have an additional factor proportional to
$N^{1/4}$. This $N$-dependence of the escape rate is reasonably well
understood, in terms of the time taken for stars to escape
\citep{2001MNRAS.325.1323B}, and is included in the more refined
models in Appendices~\ref{A} and \ref{C}.

\begin{table}
\caption{Overview of literature results for the escape rate
  $\ndot\tcrj$ in post-collapse/balanced evolution for clusters with different
  $\mu\equiv\mmax/\mmin$. Values are derived from the results of
  (1)~\citet{1961AnAp...24..369H} , (2)~\citet{1987ApJ...322..123L},
  (3)~\citet*{1991ApJ...366..455L}, (4)~\citet{1998HiA....11..591H},
  (5)~\citet{2008MNRAS.389L..28G}, (6)~\citet*{2010MNRAS.409..305L}. }
\label{tab:ndot}
\begin{tabular}{llll}
\hline
Reference & $-\ndot\tcrj$                    & $\mu$     & Comments\\\hline
   $(1)$  & 95                              &  1        & Theory\\
   $(2)$  & 76                               &  1        & Fokker-Planck\\
   $(3)$  & 200--250                         &  7        & Fokker-Planck \\
   $(4)$  & $290\left(N/10^5\right)^{\!1/4}$  & 15        & $N$-body\\
   $(5)$  & $300\left(N/10^5\right)^{\!1/4}$  & 30        & $N$-body\\
   $(6)$  & $270\left(N/10^5\right)^{\!1/4}$  & $\sim$10  & $N$-body+stellar evolution\\
\hline
\end{tabular}
{\newline Notes on the data.
  \begin{enumerate}
    \item[(2)] Based on their value of $A$ (p.128).
\item[(3)] Based on their value of $\tau_{\rm ev}$ (p.458) and a
  mass-function index $x=1$.
\item[(4)] Based on a run with $N = 65\,536$.  The dependence on $N$ has
  been assumed from \citet{2001MNRAS.325.1323B}.
\item[(5)] Based on the results of their Fig.~2. 
This study found a slightly stronger $N$-dependence; 
the $N^{1/4}$ was assumed and matched to the runs with the largest number of particles ($N=32\,768$).
\item[(6)] From a fit to the data presented in their Fig.~4
  (Lamers 2010, private communication). This study also finds a
  slightly stronger $N$-dependence; the $N^{1/4}$ was again assumed
  and matched to the runs with the largest number of particles
  ($N=131\,072$).
  \end{enumerate}
}
\end{table}

%%%%%%%%%%%%%%%%%%%%%%%%%%%%%%%%%%%%%%%%%%%%
\section{Comparison to the globular clusters in the Milky Way }
\label{sec:obs}

%___________________________________________________________________
\subsection{Model parameters}
\label{ssec:model-params}

In this section we compare the basic model of the previous section to
parameters of Milky Way globular clusters.  We convert the model
results for $N$ to $M$ assuming $\mmean=0.5$ and convert $\tcr$ to
$\rhoh$ (equation~\ref{eq:tcr}). The tidal field parameters $\tcrmax$
and $\tcrj$ (equations~\ref{eq:tcrmax} and \ref{eq:tcrj}) are
converted to $\rg$ by approximating the Milky Way potential by {that
  of} an isothermal sphere with a (constant) circular velocity of
$\vg=220\,\kms$ and using \henon's value $\rhrjmaxtext=0.145$. The
resulting formulae are given in Appendix~\ref{B}.  With these
parameters fixed, the speed of evolution is set by the value of
$\zeta$ which is chosen to be $\zeta=0.2$; this is appropriate for
clusters with a globular cluster-type mass spectrum ($\mu\simeq10$,
Section~\ref{ssec:physical}). As a consistency check, by
equation~(\ref{eq:tcrjndotn}), we derive $\ndot\tcrj\simeq-256$, which
may be expressed as (see Appendix~\ref{B})
\smallskip
\begin{equation}
\dot{M}\rg\simeq-20\,\msun\,\myr^{-1}\,\kpc .
\label{eq:mdot}
\end{equation}
The first of these is in
reasonable agreement with the evaporation rates found in both
$N$-body {and Fokker-Planck} models {with a comparable mass
function} (Table~\ref{tab:ndot}).  

From these data we can obtain the value of $\tevn$ from substitution
into the equations of Section~\ref{ssec:tev}. This then gives us the
relation between $\rhoh$, $M$ and $\rg$ at a given time (isochrones)
and the evolution of the individual parameters in time
$\rhoh(M_0,\rg,t)$, $M(M_0,\rg,t)$ (tracks).  In Appendix~\ref{B} we
provide these relations and also some of the intermediate steps and
additional relations for $\rh(M, \rg)$ and $\rh(M_0,\rg,t)$ that are
not shown here.

\begin{figure}
 \includegraphics[width=8.cm]{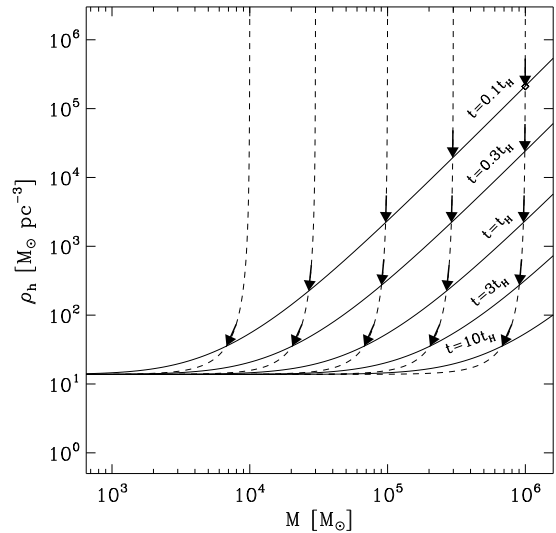}
     \caption{Isochrones (full lines) and evolutionary tracks (dashed
       lines) extracted from the model described in
       Section~\ref{sec:cycle},  applied to clusters orbiting at a
         radius of $\rg=8\,\kpc$ in an isothermal halo with circular
         velocity of $\vg=220\,\kms$. Arrows along the tracks
       indicate the direction of evolution and have a length of 0.2
       dex in time. The central isochrones has an age of a Hubble time
       $\thubble=13\,\gyr$.}
   \label{fig:tracks}
\end{figure}

In Fig.~\ref{fig:tracks} we illustrate the behaviour of the
evolutionary tracks (dashed lines) together with various isochrones
(full lines) in these more appealing quantities for clusters at
$\rg=8\,\kpc$. The isochrones span 2 dex in age centred around a
Hubble time ($\thubble=13\,\gyr$). The asymptotic behaviour of the
isochrones at the extremes is easy to understand. In the expansion
dominated phase (right-hand side of Fig.~\ref{fig:tracks}) all
clusters have evolved to the same relaxation time
(cf. equation~\ref{eq:tsimtrh}) and thus $\rhoh\propto (M/t)^2$. The
proportionality, i.e. the absolute vertical positioning of the lines,
is set by the value of $\zeta$.  In the evaporation-dominated regime
(left-hand side of Fig.~\ref{fig:tracks}) $\rhoh$ has adjusted to the
tidal density and $\rhoh\propto \rg^{-2}$.  This is because clusters
are in the homologous phase where $\rh/\rj=\rhrjmaxtext$ and therefore
$\rhoh\propto\rhoj$; {while} for an isothermal halo the Jacobi
density, $\rhoj$ scales with $\rg$ as $\rhoj\propto\rg^{-2}$
(Appendix~\ref{B}). The tracks (dashed lines) show that clusters
initially move down in this diagram, because they are expanding
without losing much mass.  Evaporation becomes more important than
expansion when $\rhoh$ approaches its minimum value and the clusters
start moving to the left. In the limit of $t\rightarrow\infty$ both
the tracks and the isochrones are horizontal lines. The speed of the
horizontal motion is set by the value of $\zeta$ and $\rg$ because {it
  follows from equation~(\ref{eq:ndottcrj}) that} $\dot{M} \propto
\zeta/\rg$.
\begin{figure*}
\center \includegraphics[width=16.cm]{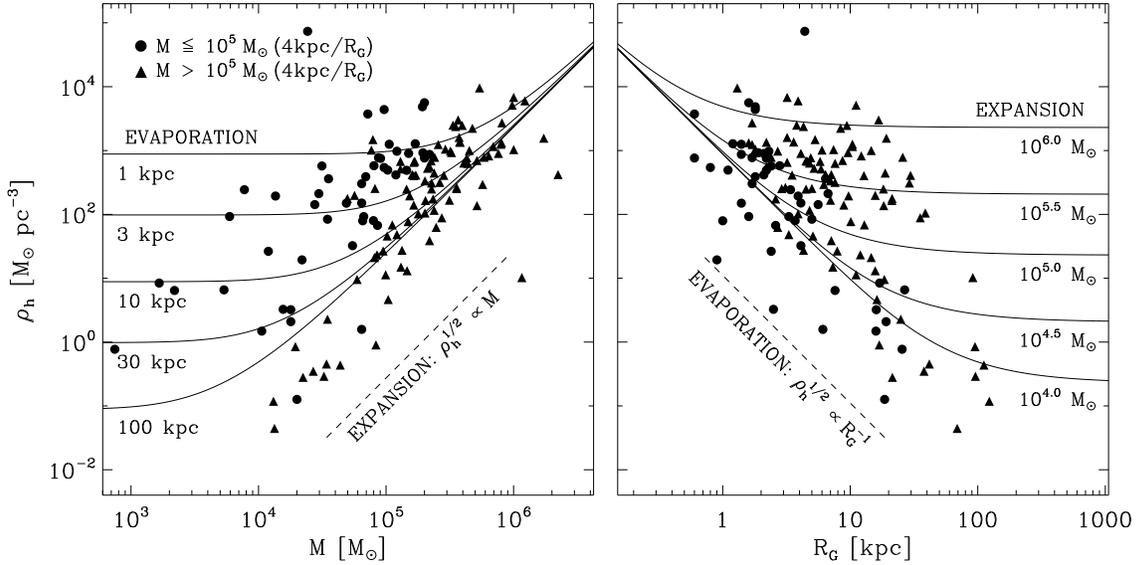}
     \caption{Isochrones based on the balanced evolution models of
       Section~\ref{sec:cycle}. In the left panel 5 isochrones for
       different $\rg$ values are shown and in the right panel 5
       isochrones for different $M$. Both panels show the 141 globular
       clusters in the \citet{1996AJ....112.1487H} catalogue from
       which $M$, $\rh$ and $\rg$ estimates can be derived. The 
       filled circles are the 48 clusters classified as being in the
       evaporation-dominated regime according to
       equation~(\ref{eq:mrg}). The traingles are the
       remaining 93 classified as being in the expansion dominated
       regime.}
   \label{fig:scaling}
\end{figure*}

%_________________________________________________________________
\subsection{The evolutionary state of Milky Way globular clusters: The relative importance of expansion and evaporation}
\label{ssec:rel-imp}
In order to see if these types of predictions are relevant for real
globular clusters we will compare our results to the globular clusters
of the Milky Way. We use the 2003 version of the
\citet{1996AJ....112.1487H} catalogue.  It contains entries for 150
globular clusters, and for 141 of them a luminosity, radius and
galacto-centric radius determination are available.  To convert
luminosity to mass we adopt a mass-to-light ratio of 2
\citep{2005ApJS..161..304M} and we multiply the projected half-light
radius by 4/3 to correct for the effect of projection
\citep{1987degc.book.....S} and get an estimate for $\rh$. Note that
by doing this we assume that light traces mass, which is not
necessarily true if the cluster is mass segregated
\citep[e.g.][]{1991ApJ...366..455L,2007MNRAS.379...93H}.
\citet{2008A&A...479..741B} study this effect in a sample of 11
globular clusters. They compare the half-light radius to the radius
containing half the number of stars and find that the latter is
typically $\sim10\%$ larger.  \citet{2007MNRAS.379...93H} finds from
$N$-body simulations that $\rh$ can be about a factor two larger than
the projected half-light radius, due to mass segregation. This implies
that for some globular clusters $\rh$ could be a factor $2/(4/3)=1.5$
higher than what we derived in Section~\ref{ssec:rel-imp}.  This might
affect clusters of different masses in different ways; this is further
discussed in Section~\ref{ssec:stellar-mass-function}. For now we
assume that light traces mass and that the correction for projection
provides a sufficiently accurate estimate of the 3-dimensional
half-mass radius $\rh$.

The first thing we determine from the data is the fraction of globular
clusters that are in the expansion dominated phase. We adopt the
definition of the boundary between these two phases as the moment
where the time derivative of $\trh$ is zero as outlined in
Section~\ref{ssec:tev}. In terms of $\tevn$ this point is when the
cluster's age is $0.4\tevn$ approximately. This means that we need to
find the parameters of clusters that at present have a life-expectancy
of $\tev\simeq1.5\thubble$, i.e.with $60\%$ of their evolution still
ahead in time.  Note that our estimates are based on an assumption
that all clusters have spent the last $\thubble$ at the $\rg$ where
they are now and that the potential of the Milky Way halo has been
constant.  This is not generally expected to be true and the
consequences of this assumption are discussed in
Sections~\ref{ssec:cluster-orbits} and
\ref{ssec:static-potential}. The exercise is, therefore, merely
intended as a first order approximation to see if it is at all
reasonable to expect that globular clusters are still in the expansion
phase.

With the adopted model parameters (Section~\ref{ssec:model-params}) and
$\thubble=13\,\gyr$ we find that clusters with
\begin{equation}
  M\gtrsim 10^5\,\msun\left(\frac{4\,\kpc}{\rg}\right)\label{eq:mrg}
\end{equation}
\noindent
should still be in the expansion-dominated
phase. (Equation~\ref{eq:mdot} is the easiest route to this result.)
This relation is satisfied by 93 of the 141 clusters.  It follows that
the remaining 48 clusters are in the evaporation-dominated phase.
This perhaps surprising result has some interesting consequences. The
most important one is that the present day densities of the majority
of the globular clusters follow (roughly) from the self-similar
expansion model for isolated clusters: $\rhoh\propto M^2$. Here we
have assumed that all clusters have the same age. This scaling
relation should be universal, because it is driven by internal 2-body
relaxation; therefore a similar scaling, with the same
proportionality, should also hold for extra-galactic clusters.
Moreover, in extra-galactic cluster samples the fraction of clusters
in the expansion-dominated phase is probably larger; they are easier
to detect because they have (on average) higher mass
(equation~\ref{eq:mrg}).

\begin{figure}
\center\includegraphics[width=8.75cm]{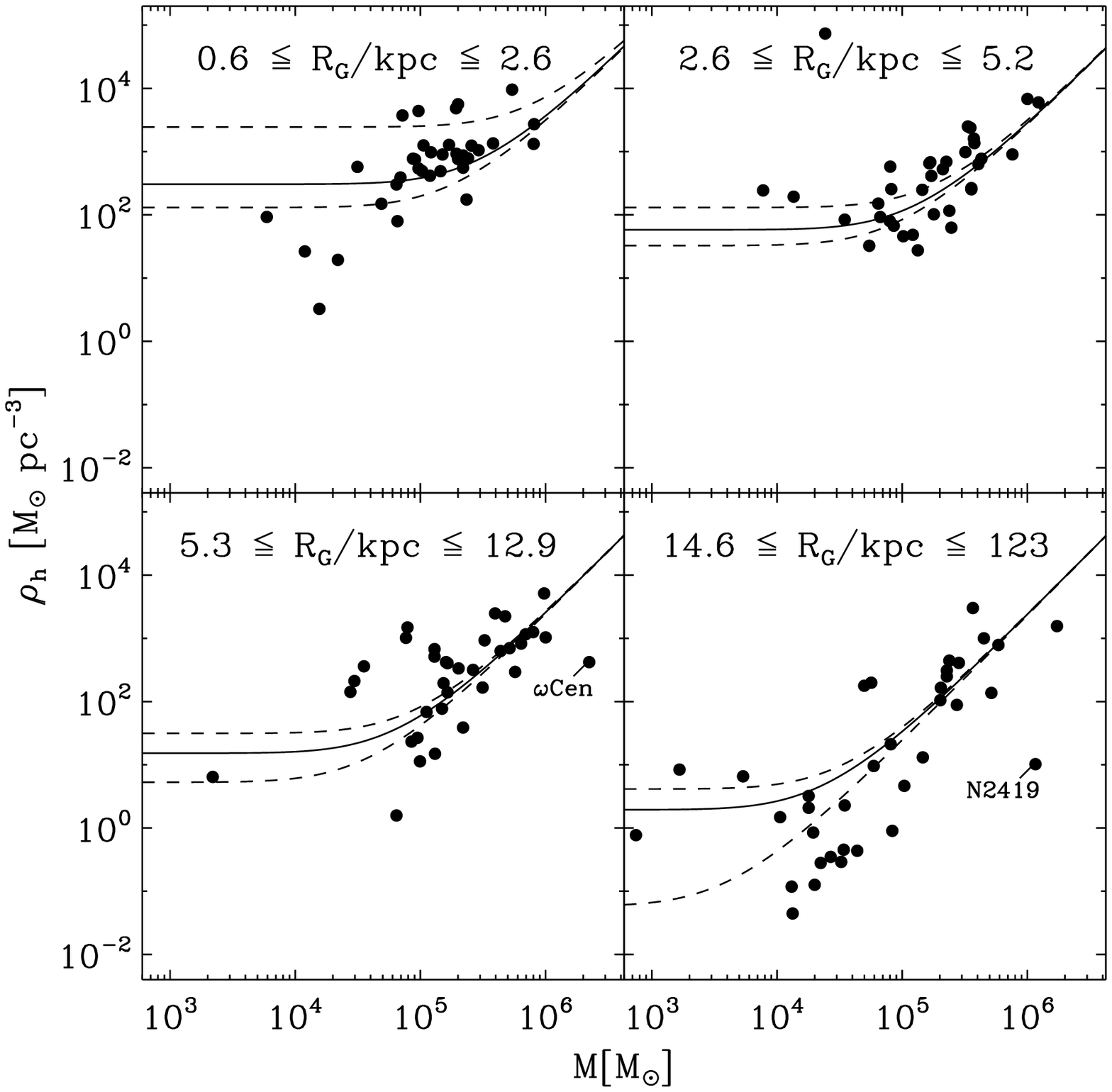} 
     \caption{Similar to left panel of Fig.~\ref{fig:tracks}, but now
       the data is sorted in $\rg$ and sub-divided in four (roughly)
       equal samples. The range in $\rg$ values is indicated in each
       panel. The full lines show isochrone predictions for the median $\rg$
       values in the samples and the dashed lines show the isochrones
       for the minimum and maximum $\rg$ values. }
   \label{fig:rho_m_data}
\end{figure}

The prediction that a $\rhoh^{1/2}\propto M$ scaling must hold for the
majority of the Milky Way globular clusters is one of the main results
of this paper.  More detailed discussion on its implications and
consequences are {postponed to} Section~\ref{sec:discussion}. First we
show in Section~\ref{ssec:isochrones} that this scaling is indeed
found for globular clusters and that the proportionality agrees with
the parameters of energy conduction derived in
Section~\ref{sec:cycle}.

%___________________________________________________________________
\subsection{Isochrones and Milky Way globular clusters}
\label{ssec:isochrones}

Because all globular clusters have roughly the same age we focus on
isochrones with an age of $\thubble\simeq13\,\gyr$, rather than the
evolutionary tracks. In Fig.~\ref{fig:scaling} we show the isochrones
in $\rhoh\,\vs M$ (left) and $\rhoh\,\vs\rg$ (right) diagrams together
with the 141 globular clusters for which data are available in the
Harris catalogue. The clusters that are in the evaporation-dominated
phase are shown as filled circles and the clusters that are
still expanding are shown as triangles.

In the left panel isochrones for clusters at different $\rg$ between
1\,kpc and 100\,kpc are shown. The isochrones roughly encompass the
data. The 100\,kpc isochrone clearly shows the asymptotic
$\rhoh^{1/2}\propto M$ behaviour following from expansion, which roughly
follows the lower envelope of data points. In the outer halo the tidal
field is so weak that all clusters with $M\gtrsim10^4\,\msun$ {have
  not expanded up to their tidal boundary yet}.  In the right panel
the densities are shown as a function of $\rg$ together with five
isochrones for different masses. As was the case for the isochrones in
the $\rhoh\,\vs\, M$ plot, these isochrones also roughly encompass the
data.  The asymptotic behaviour of the isochrones in both diagrams is
given by labels in the two diagrams.

\begin{figure}
\center \includegraphics[width=8.75cm]{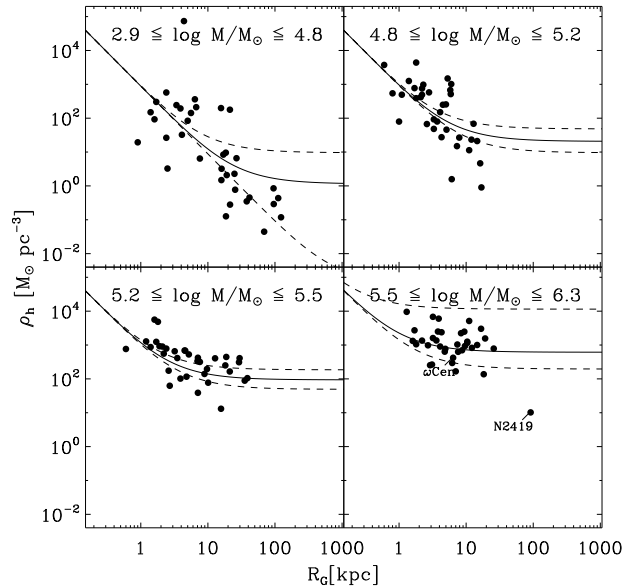} 
     \caption{Similar to right panel of Fig.~\ref{fig:tracks}, but now
       the data is sorted in $M$ and sub-divided in four (roughly)
       equal samples. The range in $M$ values is indicated in each
       panel. The full lines show isochrones predictions for the
       median values of $M$ in the samples and the dashed lines show the
       isochrones for the minimum and maximum values. }
   \label{fig:rho_rg_data}
\end{figure}

To further illustrate the comparison between data and theory we
subdivided the Milky Way sample in four samples containing roughly equal
numbers, but sorted in $M$ and $\rg$. We then compute the isochrones
based on the median, the minimum and maximum value in each sample. The
results for different $\rg$ bins and different $M$ bins are shown in
Fig.~\ref{fig:rho_m_data} and Fig.~\ref{fig:rho_rg_data},
respectively. The general relation between $M$, $\rhoh$ and $\rg$ is
nicely matched by the model.

Our model is based on the assumption of {\it balanced} evolution
(Section~\ref{sec:intro}), which led to the conclusion that
$\trh\lesssim 0.3\thubble$ (Section~\ref{ssec:tev},
equation~\ref{eq:tsimtrh}).  A logical conclusion of our results,
then, is that the evolution of objects with $\trh\gtrsim 0.3\thubble$
is unbalanced, presumably implying that they have formed with a long
$\trh$ \citep[this scenario was explored using numerical simulations
  by][]{2010MNRAS.408.2353H,2011MNRAS.411.1989Z}. If we draw a hard
line at this limit then we find 23 globular clusters for which this
holds. However, {many} of these clusters are fairly close to the
$\trh=0.3\thubble$ line (see left panel of Fig.~\ref{fig:scaling}).
The majority of the points which are substantially below the lowest
isochrone have a low mass, and given the uncertainties in radius, 
  mass and distance estimates for these objects we should not exclude
the possibility that their positioning in the diagram is consistent
with the model within measurement uncertainties. If these objects are
in balanced evolution, they have undergone a lot of dynamical
evolution, and could therefore be mass segregated, and perhaps
estimates for $N$, $\mmean$ and $\rh$ are systematically affected by
this.  In Section~\ref{sec:discussion} we further discuss this issue
of the clusters with $\trh\gtrsim0.3\thubble$.  Obvious exceptions
(i.e. high-mass clusters well below the lowest isochrone) are NGC~5139
($\omega$Cen) and NGC~2419 with $\trh\simeq16\,\gyr$ and
$\trh\simeq54\,\gyr$, respectively. These are the only two objects in
the Harris catalogues with $\trh\gtrsim\thubble$.  Due to their high
mass it is unlikely that the data of these objects suffer from
measurement uncertainties and it is more likely that these objects
have properties resembling their initial conditions \citep[besides
  some adiabatic expansion because of mass-loss due to stellar
  evolution perhaps,][]{1980ApJ...235..986H, 2010MNRAS.408L..16G}.

%%%%%%%%%%%%%%%%%%%%%%%%%%%%%%%%%%%%%%%%%%%%
\section{Summary and discussion}
\label{sec:discussion}
In this study we provide a description of the evolution of mass and
half-mass density (or radius) of (globular) clusters. The two
self-similar solutions provided by \citet{1965AnAp...28...62H} and
\citet{1961AnAp...24..369H} for clusters evolving in isolation and in
a tidal field, respectively, are combined into a pair of Unified
Equations of Evolution (UEE), which yield analytically continuous
evolutionary tracks and isochrones. The key ingredient is that
clusters are assumed to conduct a constant fraction of their total
energy per half-mass relaxation time-scale.  This is assumption is
justified by the fact that both \henon\ models, i.e. those that
describe the extremes \citep{1961AnAp...24..369H,1965AnAp...28...62H},
have a very similar $(\dot{E}/\eabs)\trh\simeq0.08$ (see
Section~\ref{ssec:eflux}).  The central energy source could be assumed
to be hard binaries in the centre of the cluster, but the details of
the energy production are not of interest to us, since we assume that
the evolution is {\it balanced}, i.e. that the amount of energy
produced by the core is determined by how much energy can be conducted
through the half-mass boundary. This important view on the evolution
of collisional gravitational systems was introduced by
\citet{1975IAUS...69..133H} and can be compared to Eddington's view on
stellar energy sources. The bottleneck in the energy emission from a
star is the radiative transfer through the envelope. The luminosity of
the star is thus set by how much radiation can be transported through
the envelope and the nuclear fusion rate in the core adjusts to this
(see \citealt{1968MNRAS.138..495L} and \citealt{1983MNRAS.205..913I}
for more details on this analogy).

The efficiency of energy transport in a cluster depends on the mass
spectrum of the stars. For a mass function typical of a globular
cluster it is about a factor of $\sim2.5$ higher than for the equal
mass clusters consider{ed} by \henon. It is this efficiency that
determines the positioning of the isochrones and the speed with which
clusters move along the tracks. The solutions are compared to the
parameters of Milky Way globular clusters and reasonable agreement is
found for almost all well-observed clusters.  We interpret this
agreement as evidence for the fact that the evolution of almost all
globular clusters is balanced, meaning that there is a central energy
source that supplies the energy needed by 2-body relaxation.
Exceptions are NGC~2419 and $\omega$Cen because their relaxation
time-scales are much longer than a Hubble time. The evolution of these
clusters is unbalanced. We find that roughly two-thirds of the Milky
Way globular clusters are in the expansion dominated phase, i.e. their
evolution is not yet seriously affected by the Galactic tidal
field. These clusters align along lines of constant relaxation time
implying that $\rhoh\propto M^2$.  This idea is further supported by
the properties of intra-cluster globular clusters in the Virgo
Cluster, which have no tidal boundary. These have masses of
$M\simeq10^5\,\msun$ and radii of a few parsecs
\citep{2007ApJ...654..835W}, comparable to Milky Way globular clusters.

%___________________________________________________________________
\subsection{Are all globular clusters in post-collapse evolution?}
\label{ssec:pcc}
We have referred to balanced evolution when we talked about clusters
with a central energy source.  For clusters without primordial
binaries and stellar evolution this is generally referred to as
post-collapse evolution, the energy being provided by binaries formed
in three-body interactions. It is interesting to enquire whether these
concepts are also related in more realistic models.

At first sight the proportion of Galactic globular clusters which are
classified as post-collapse objects is quite inconsistent with our
conclusion that the evolution of almost all globular clusters is
balanced.  Only $\sim20$\% of the globular clusters possess the steep
cusp in their surface-brightness profile \citep{1986ApJ...305L..61D}
that is predicted by models of post-collapse evolution.  We should
bear in mind, however, that such predictions are based on studies of
equal-mass systems.  In a system with an evolved stellar population,
however, it is possible that the distribution of the remnant
population is steeply cusped, and that the central distribution of
those stars which dominate the surface-brightness exhibits only a
shallow cusp.  This has been shown in many studies since at least the
work on M15 by \citet{1977ApJ...218L.109I};
\citet{1984ApJ...280..298G} gives a general and simple theoretical
explanation.

Several independent arguments support the view that the
surface-brightness profile is not a reliable guide to the post- or
pre-collapse status of a cluster.  \citet{1984ApJ...277L..45C} argued
that globular clusters with central relaxation times less than
$10^8\,$yr have already undergone core-collapse, which is close to
half of their sample of 146 Galactic globular clusters. Next,
\citet{2007ApJ...656L..65D}, on the basis of a surprising correlation
between the concentration of a cluster and the slope of the main
sequence mass function, argued that the number of post-collapse
globular clusters has been seriously underestimated.  Finally,
specific examples are being revealed by detailed modelling of
individual clusters. A good example is the cluster M4, whose
surface-brightness profile is well described by a classical King
model, but dynamical modelling suggests that it has already undergone
core-collapse \citep{2008MNRAS.389.1858H}.

While these arguments reduce the gap between the proportion of
post-collapse clusters, on the one hand, and the proportion exhibiting
balanced evolution on the other, they do not imply that the two
concepts are identical.  In another modelling exercise
\citet{2011MNRAS.410.2698G} argue that 47 Tuc is far from core
collapse, and yet the model exhibits what we would call balanced
evolution, the required energy being supplied by mass loss from
stellar evolution and, to a lesser extent, heating by primordial
binaries.  In this model, what is called `core collapse' results from
the long-term exhaustion of these energy sources.

Models of globular clusters show an early rapid collapse of the core
radius (e.g. Fig.\ref{fig:mc-model}) caused by mass segregation. It is
also exhibited by the model of 47 Tuc which we have just been
discussing.  This phenomenon is the closest analogue of the classical
core collapse exhibited by single-component models, and perhaps the
main lesson of this discussion is that we need a better, agreed
definition of what we mean by core collapse.

%____________________________________________
\subsection{Cluster orbits}
\label{ssec:cluster-orbits}
In our model we have made the implicit assumption that cluster orbits
are circular in the Milky Way potential.  This is of course not true.
\citet{1999AJ....117.1792D} determined the orbits of 38 globular
clusters and they find a median eccentricity of $\ecc\simeq0.65$.
The question is to what extent this affects our interpretation.

Clusters that are in the expansion-dominated phase are not affected by
the tide, and because we have shown that this is the case for the
majority of the globular clusters in the Milky Way, the details of the
shape of the orbits should not affect the result that the empirically
established scaling of $\rhoh^{1/2}\propto M$ is a result of
expansion.  However, our estimate of the fraction of clusters in the
expansion dominated phase was based on the assumption of circular
orbits.  This fraction could very well be smaller because a cluster
that is in the expansion phase based on its density and its current
galacto-centric position could be in the mass loss dominated phase at
peri-centre ($\rp$) when we allow for the possibility that $\ecc>0$.

Despite what is commonly assumed, conditions at $\rp$ are not
relevant, either for the structure of the cluster or for the evaporation
rate. On the first point, \citet{2010MNRAS.407.2241K} showed that the
density profile of the bulk of the stars adjusts to the mean tidal
radius along the orbit, rather than the tidal radius at
perigalacticon.  On the second point, again through use of $N$-body
simulations, it has also been shown that the escape rate of a cluster
on an oval galactic orbit is close  to  that of a cluster on a circular
orbit with a radius intermediate between $\rp$ and the apogalactic
distance \citep{2003MNRAS.340..227B}.  Though it has not been shown
that the galactocentric radius determining the evaporation  coincides
with that characterising the structure, the systematic error in
fitting the evolution of a cluster to its present radius must be
considerably smaller than would be the case if structure and mass loss
were set at $\rp$.  Nevertheless, whatever the magnitude of this
effect, it does not help to explain those clusters which, in
Fig.\ref{fig:scaling} (left panel), lie below the envelope of the
isochrones.

%____________________________________________
\subsection{Static galaxy potential}
\label{ssec:static-potential}

We have assumed that clusters have been evolving for a Hubble time at
the galactocentric position where they are observed now.  This is an
oversimplification, because in the hierarchical merging scenario
clusters can form in dwarf galaxies that are later accreted in larger
haloes \citep{2005ApJ...623..650K, 2008ApJ...689..919P,
  2010ApJ...717L..11M}. Most of the galaxies are in place at a
redshift $z\simeq2-3$, hence the assumption of a static galactic
potential only affects the clusters that were brought into haloes in
late accretion events. In fact, there is the potential of using the
densities of clusters to say something about the accretion
history. Imagine a cluster in a dwarf galaxy with a strong tidal
field. Suppose that in this host it has already reached the
evaporation-dominated phase, i.e. its density has adjusted to the
tidal density in the dwarf galaxy \citep[which is the case for the
  clusters associated with the Sagittarius dwarf
  galaxy,][]{2009MNRAS.399.1275P}. If the dwarf galaxy is accreted in
a larger host, such as the Milky Way, the new tidal field that the
cluster experiences is in most cases weaker than in the previous
host. The cluster will start expanding again, but its density is
higher than the prediction based on a Hubble time of evolution at that
$\rg$. A target to which this might apply is Palomar~1 \citep[see the
  discussion in][]{2010MNRAS.408L..66N}.

%____________________________________________
\subsection{Model validity at younger ages}
Throughout this study we have assumed a constant value of
$\zeta\simeq0.2$ for all ages.  \citet{2010MNRAS.408L..16G} found
values for $\zeta$ as large as $2$ for clusters with a full mass
spectrum ($\mu=1000$), which would be appropriate for very young star
clusters ($\lesssim3$\,Myr).  We have also assumed that the evolution
is balanced at all ages.  Therefore we need to bear in mind that the
results presented here should not be interpreted in a literal way for
clusters with ages much younger than a Hubble time. The results
presented here underestimate the speed of balanced evolution for
younger clusters. For clusters that have not yet reached the balanced
evolution phase our results overestimate the speed of evolution.  The
expansion for clusters with young ages is considered in more detail by
\citet{2010MNRAS.408L..16G} and we refer to that paper for an analytic
description of the radius as a function of initial $\trh$ considering
the effect of the changing stellar mass function and the transition
from un-balanced to balanced evolution.  A time-dependent functional
form for $\zeta$ could in principle be found from the time-dependent
parameter $\chi_t$ of \citet{2010MNRAS.408L..16G}.  We refrain from
doing this, however, because $\chi_t$ depends on the details of the
stellar mass function, which is affected by both stellar evolution and
by the loss of low-mass stars over the tidal boundary.  The latter was
not taken into account in the models of
\citep[][]{2010MNRAS.408L..16G}. We therefore leave these details of
the full evolution for a future study.

%____________________________________________
\subsection{The stellar mass function}
\label{ssec:stellar-mass-function}
In Section~\ref{ssec:isochrones} we found that there are a number of
low-mass clusters with $0.3\thubble\lesssim\trh\lesssim \thubble$,
i.e. their relaxation times seem inconsistent with the predictions for
expansion with $\zeta=0.2$.  Several Palomar clusters are in this
group.  We now consider a number of factors which might account for
such cases.

Because these clusters are faint, they could be close to final
dissolution and might have stellar mass functions that are depleted in
low mass stars \citep{2003MNRAS.340..227B}.  We assumed
(Section~\ref{ssec:physical}) that the value of $\zeta$, which
controls the rate of evolution, depends only on the range of the mass
function, i.e. the ratio $\mu$ of the maximum to minimum mass, and
this does not depend on the slope of the mass function.
\citet{2004ApJ...604..632G} showed that for the rate of core collapse
it is not $\mu=\mmax/\mmin$ that sets the speed of evolution, but
rather $\mmax/\mmean$, where $\mmean$ is the mean mass.  If the same
consideration applies to balanced evolution, then for flatter mass
spectra, i.e. lower $\mmax/\mmean$, we would expect a slower dynamical
evolution, i.e. smaller $\zeta$.  By equation~(\ref{eq:tev}) this
increases the ratio $\tevn/\tcrmax$.  By use of
equation~(\ref{eq:trhmax}) we see from equation~(\ref{eq:tsimtrh})
that the effect is to reduce the ratio $\trh/t$ in the expansion
phase.  Thus for a given mass, clusters should be denser at a given
age.  On the other hand Fig.\ref{fig:scaling} (left panel) shows that
the clusters of concern are less dense than expected.  We conclude,
therefore, that a flatter mass function can not be responsible for an
additional decrease of the densities of these low mass clusters.

Another consideration is the $N$-dependence in the escape rate which
is implicit in the $N$-body results quoted in Table
\ref{tab:ndot}. This affects the expansion rate $\chi$ of clusters
that are well into the evaporation-dominated regime (Appendix~\ref{A})
and therefore does not help us to understand the discrepant objects.

We propose two alternative speculative explanations.  Firstly, it
could be that these clusters had a high retention fraction of black
holes and neutron stars.  If the most massive stars are black holes
with $m\simeq10-20\,\msun$, the ratio $\mu$ would be higher and the
dynamical evolution would be faster. As mentioned in
Section~\ref{ssec:physical}, \citet{2010MNRAS.408L..16G} show that
$\zeta\propto\mu^{1/2}$ approximately. For a mass function between
$0.1\,\msun$ and $1\,\msun$ we have adopted $\zeta\simeq0.2$.  For a
cluster with a few additional black holes with masses of $10\,\msun$
we would have $\zeta$ approximately a factor of $\sim3$ higher. This
would help to make all of these clusters consistent with balanced
evolution.

Secondly, we have assumed that the mass-to-light ratio is the same for
all clusters.  But if the cluster is depleted in low-mass stars we
surely overestimate $N$ and underestimate $\mmean$ by using a
canonical mass-to-light ratio.  Because $\trh \propto
(N/\mmean)^{1/2}$ we thus overestimate $\trh$.  Thus if, according to
our estimates, a cluster has a half-mass relaxation time-scale
$>0.3\thubble$, it may be that its actual $\trh$ is more consistent
with our theory.  It would be interesting to consider these effects in
more detail with numerical simulations.

Lastly, estimates of $\trh$ are very sensitive to uncertainties
  in the distance to the cluster, $D$. The radius in pc depends
  linearly on $D$ and the mass in solar masses quadratically, such
  that $\trh\propto D^{5/2}$. Because $\rhoh\propto D^{-1}$ and
  $M\propto D^2$ the scatter of data points due to uncertainties in
  $D$ is orthogonal to lines of constant relaxation time (left panel
  of Fig.~\ref{fig:scaling} and Fig.~\ref{fig:rho_m_data}). 

%____________________________________________
\subsection{Comparison to other work}
The relation of globular clusters to lines of constant relaxation time
($\rhoh^{1/2}\propto M$) has been noted for Milky Way globular
clusters and for extra-galactic globular clusters
\citep[e.g.][]{1997ApJ...474..223G,2007AJ....133.2764B,
  2008ApJ...679.1272M,2009MNRAS.392..879G}.  It is usually given a
different interpretation, however.  Because for homologous clusters
the life-expectancy is a constant number of relaxation times ($1/\xi$)
the lines of constant $\trh$ have been interpreted as {\it boundaries}
to the expected distribution of surviving clusters.  The first to
provide this interpretation were probably \citet{1977MNRAS.181P..37F}
who showed the dependence on mass and radius of various disruptive
agents.  Given that the time-scales for relaxation driven evaporation
(of homologous clusters), disruption by tidal shocks and dynamical
friction all depend in a different way on cluster mass and radius one
can construct a `survival' or `vital' triangle in an $\rh\vs M$
diagram \citep[e.g.][]{1997ApJ...474..223G}, or alternatively
$\rhoh\vs M$. Clusters within this triangle are likely to survive for
another Hubble time.

In this paper we have given a different interpretation to the relation
between cluster data and lines of constant $\trh$.  Because the
majority of the clusters have not reached the homologous (tidally
limited) evaporation-dominated phase yet, the alignment of the
globular cluster data for these clusters with the $\trh\simeq
0.3\thubble$ relation results from expansion driven by 2-body
relaxation {when evolution is balanced}.  There would be no globular
clusters with shorter relaxation times if there was no tidal field
that can stop the expansion. In reality clusters can not expand
indefinitely, and the relaxation time starts decreasing once
evaporation starts to dominate over expansion. This is why there are
clusters with $\trh<0.3\thubble$.  Note that this is just the opposite
of the usual interpretation, which implies that only clusters with
{\it longer} relaxation times should be observed!

Because $\trh$ decreases linearly in time in the homologous phase,
there should be a uniform distribution of $\trh$ values between $0$
and $\sim0.3\thubble$. This was proposed by
\citet{1995ApJ...443..109L} and they show that this indeed roughly
holds. Since in the homologous phase $\trh\propto M$, there should
also be a uniform distribution of $M$ at low mass, or $\dr N/\dr
M\simeq\,$const \citep{1961AnAp...24..369H, 2001ApJ...561..751F},
which is indeed found in many galaxies \citep{2007ApJS..171..101J}.

\citet{2008MNRAS.389..889K} recently presented results of $N$-body
simulations of initial compact clusters in a tidal field. They refer
to the phase where the half-mass radius has adjusted to the Jacobi
radius as `main sequence' evolution.  Using again the analogy with
stars we could, however, refer to the entire balanced evolution,
i.e. expansion and evaporation, as main sequence evolution.  The
initial unbalanced evolution without a central energy source would
then be the `pre-main sequence evolution'. On the other hand perhaps
the analogy should not be pushed too far; for
\citet{1968MNRAS.138..495L} the end of core collapse signalled the
start of the red giant phase!

\citet{2010MNRAS.401.1832B} identified two populations of clusters
based on the ratio of the half-mass radius over the Jacobi radius
($\rh/\rj$).  About half of the clusters beyond $\gtrsim8\,\kpc$ is
strongly under-filling their tidal radius (i.e. $\rh<<\rj$), while the
other half have $0.1\lesssim\rh/\rj\lesssim0.3$.  They find a strong
correlation between mass and the membership of these groups: the
under-filling group contains significantly more-massive clusters than
the `tidally filling' group.  This separation is because low-mass
clusters have already expanded towards their tidal boundary and all
have a similar ratio $\rh/\rj$.  Their more massive counterparts are
still expanding towards their tidal boundary (see right panel of
Fig.~\ref{fig:scaling}).  The densities of clusters in the
expansion-dominated regime within $\rg\lesssim8\,\kpc$ are more
similar to those of the clusters that are already in the
evaporation-dominated regime (Fig.~\ref{fig:scaling}) and it is,
therefore, harder to make a distinction between under-filling and
filling based on the empirically determined ratio $\rh/\rj$.

%____________________________________________
\subsection{Implications for the use of globular clusters as standard rulers for distance estimation}
Globular cluster properties have been used as standard candles for
distance estimates. Traditionally the peak of the globular cluster
luminosity function (GCLF), because of its near universality, is used
for this \citep[e.g.][]{1991ARA&A..29..543H}. However, even {\it if}
the initial mass function of globular clusters is universal, the
detailed shape of the GCLF will be environment dependent after a
Hubble time of dynamical evolution
\citep{1997MNRAS.289..898V,1997ApJ...487..667O,2010ApJ...717..603V}. \citet{2005ApJ...634.1002J}
suggested that the radius distribution can be used as a standard
ruler. Alternatively, we suggest to use a combination of the two as a
distance ruler; namely the relation between apparent luminosity and
radius. This can be done because for the majority of clusters the
mass-radius relation is independent of environment and second order
effects such as the stellar mass function can in principle be taken
into account \citep[see also][]{2004ApJ...613L.117J}.

%____________________________________________
\subsection{Relation to fundamental plane relations}
With all the theory in place we are able to test if empirically
established `fundamental-plane relations'
\citep[e.g.][]{1995ApJ...438L..29D, 1997AJ....114.1365B,
  2000ApJ...539..618M} can be attributed to balanced evolution.
Because we do not make predictions for the central velocity
dispersion, we will focus here on the mass-radius relation.

It is often stated that the radii of globular clusters do not
correlate with their mass \citep[e.g.][]{1991ApJ...375..594V}.  This
statement is perhaps a slight oversimplification, because there may be
a negative correlation, especially for clusters in the outer halo
\citep{2004MNRAS.354..713V, 2010MNRAS.401.1832B}.
\citet{2000ApJ...539..618M} showed that, if a correction is made for
the galacto-centric radius, the cluster half-mass radius is
approximately constant, i.e. independent of mass:
$\rh/\rg^{0.4}\sim\,$constant.

Strictly speaking the theory presented in this paper can not explain a
radius that is perfectly independent of mass. The relation between
mass and half-mass radius depends on the regime the cluster is in. For
clusters in the expansion dominate{d} regime we expect a constant
relaxation time, i.e. $\rh\propto M^{-1/3}$ independent of $\rg$,
while for clusters in the mass loss regime we expect a stronger
dependence on $\rg$ {than on $M$}: $\rh\propto M^{1/3}\rg^{2/3}$ (or
$\rh\propto M^{1/6}\rg^{2/3}$ when we use a slightly different escape
criteri{on}, see Appendix~\ref{A}).
\footnote{The correlation between $\rh$ and $M$ becomes weaker still
  if we take into account the Coulomb logarithm.  If we approximate
  $\ln\Lambda$ by a weak power of $M$: $\ln\Lambda\propto M^{0.15}$
  then we find in the expansion dominated regime that $\rh\propto
  \ln\Lambda^{2/3}M^{-1/3}\simeq M^{-0.23}$ and in the evaporation
  dominate regime $\rh\propto M^{1/3}/\ln\Lambda^{2/3}\simeq
  M^{0.23}$ (or $\rh\propto M^{1/6}/\ln\Lambda^{2/3}\simeq M^{0.07}$
  for $x=3/4$, see Appendix~\ref{A}).}.

If we split the Milky Way globular cluster sample, using the boundary
between these two extremes as described in Section~\ref{ssec:rel-imp},
then we can search for these different (bivariate) relations in the
data and see if they indeed hold. We use a straightforward linear
regression fit to find how $\log\rh$ depends on $\log M$ and $\log
\rg$. We exclude $\omega$Cen and NGC~2419 because the balanced
evolution does not apply to these objects. We also exclude NGC~6540,
with a density of $\rhoh\simeq7\times10^4\,\msunpc$, because it is an
extreme outlier in all diagrams.  From a bivariate fit to the 47
clusters in the evaporation-dominated regime we find
\begin{eqnarray}
\log(\rh/\pc)&=&-0.16+(0.11\pm0.08)\log(M/\msun)\nonumber\\
   &&+(0.44\pm0.11)\log(\rg/\kpc).
\end{eqnarray}
This shows a weak positive correlation with $M$ close to the predicted
index of $0.17(0.33)$ for $x=3/4(1)$ (see Appendices~\ref{A} and \ref{C}  for the
significance of $x$) and a relatively strong dependence on $\rg$, in
rather poorer agreement with the predicted index of $0.67$.  On the
other hand we find for the 91 clusters in the expansion dominated
regime
\begin{eqnarray}
\log(\rh/\pc)&=&1.82-(0.25\pm0.05)\log(M/\msun)\nonumber\\
   &&+(0.27\pm0.05)\log(\rg/\kpc).
\end{eqnarray}
This is consistent with a constant relaxation time ($\rh\propto
M^{-1/3}$, equation~\ref{eq:trh_spitzer_hart}), but there is still
some dependence on $\rg$, whereas pure expansion would not give rise
to any $\rg$ dependence.

The balanced evolution we consider here is thus responsible for a
slightly complicated correlation between radius and mass.  The theory
predicts that the radius has a slight positive correlation with mass
for $M\le 10^5\,\msun(4\,\kpc/\rg)$, while for more massive clusters
the correlation is negative and consistent with a constant relaxation
time relation.  This subtle behaviour of the mass-radius relation is
indeed found {(at least qualitatively)} for the Milky Way globular
clusters{, as we have just seen}. An approximately mass-independent
radius (at a given $\rg$) is thus only true to first approximation for
the population as a whole.  The strong correlation of $\rh$ with $\rg$
for the cluster population as a whole does not seem consistent with
our finding that the majority of the clusters are still expanding.
However, we note that the scaling $\rh\propto\rg^{0.4}$ is very
sensitive to the handful of very low density clusters in the outer
halo ($\gtrsim50\,\rg$).

%%%%%%%%%%%%%%%%%%%%%%%%%%%%%%%%%%%%%%%%%%%%
\section*{Acknowledgement}
MG acknowledges the Royal Society for financial support and the School
of Mathematics of the University of Edinburgh for several pleasant
visits. The authors thank Florent Renaud for providing translations
into English of the two seminal papers of Michel \henon\ and for
various interesting discussions on the effect of tides on star
clusters.  The authors also thank the referee, Donald Lynden-Bell, for
interesting suggestions and comments on the manuscript.
   
\bibliographystyle{mn2e}

\begin{thebibliography}{}

\bibitem[\protect\citeauthoryear{{Barmby}, {McLaughlin}, {Harris}, {Harris} \&
  {Forbes}}{{Barmby} et~al.}{2007}]{2007AJ....133.2764B}
{Barmby} P.,  {McLaughlin} D.~E.,  {Harris} W.~E.,  {Harris} G.~L.~H.,
  {Forbes} D.~A.,  2007, \aj, 133, 2764

\bibitem[\protect\citeauthoryear{{Baumgardt}}{{Baumgardt}}{2001}]{2001MNRAS.32%
5.1323B}
{Baumgardt} H.,  2001, \mnras, 325, 1323

\bibitem[\protect\citeauthoryear{{Baumgardt}, {Hut} \& {Heggie}}{{Baumgardt}
  et~al.}{2002}]{2002MNRAS.336.1069B}
{Baumgardt} H.,  {Hut} P.,    {Heggie} D.~C.,  2002, \mnras, 336, 1069

\bibitem[\protect\citeauthoryear{{Baumgardt} \& {Makino}}{{Baumgardt} \&
  {Makino}}{2003}]{2003MNRAS.340..227B}
{Baumgardt} H.,  {Makino} J.,  2003, \mnras, 340, 227

\bibitem[\protect\citeauthoryear{{Baumgardt}, {Makino} \&
  {Ebisuzaki}}{{Baumgardt} et~al.}{2004}]{2004ApJ...613.1143B}
{Baumgardt} H.,  {Makino} J.,    {Ebisuzaki} T.,  2004, \apj, 613, 1143

\bibitem[\protect\citeauthoryear{{Baumgardt}, {Parmentier}, {Gieles} \&
  {Vesperini}}{{Baumgardt} et~al.}{2010}]{2010MNRAS.401.1832B}
{Baumgardt} H.,  {Parmentier} G.,  {Gieles} M.,    {Vesperini} E.,  2010,
  \mnras, 401, 1832

\bibitem[\protect\citeauthoryear{{Bonatto} \& {Bica}}{{Bonatto} \&
  {Bica}}{2008}]{2008A&A...479..741B}
{Bonatto} C.,  {Bica} E.,  2008, \aap, 479, 741

\bibitem[\protect\citeauthoryear{{Burstein}, {Bender}, {Faber} \&
  {Nolthenius}}{{Burstein} et~al.}{1997}]{1997AJ....114.1365B}
{Burstein} D.,  {Bender} R.,  {Faber} S.,    {Nolthenius} R.,  1997, \aj, 114,
  1365

\bibitem[\protect\citeauthoryear{{Cohn}}{{Cohn}}{1979}]{1979ApJ...234.1036C}
{Cohn} H.,  1979, \apj, 234, 1036

\bibitem[\protect\citeauthoryear{{Cohn} \& {Hut}}{{Cohn} \&
  {Hut}}{1984}]{1984ApJ...277L..45C}
{Cohn} H.,  {Hut} P.,  1984, \apjl, 277, L45

\bibitem[\protect\citeauthoryear{{De Marchi}, {Paresce} \& {Pulone}}{{De
  Marchi} et~al.}{2007}]{2007ApJ...656L..65D}
{De Marchi} G.,  {Paresce} F.,    {Pulone} L.,  2007, \apjl, 656, L65

\bibitem[\protect\citeauthoryear{{Dinescu}, {Girard} \& {van Altena}}{{Dinescu}
  et~al.}{1999}]{1999AJ....117.1792D}
{Dinescu} D.~I.,  {Girard} T.~M.,    {van Altena} W.~F.,  1999, \aj, 117, 1792

\bibitem[\protect\citeauthoryear{{Djorgovski}}{{Djorgovski}}{1995}]{1995ApJ...%
438L..29D}
{Djorgovski} S.,  1995, \apjl, 438, L29

\bibitem[\protect\citeauthoryear{{Djorgovski} \& {King}}{{Djorgovski} \&
  {King}}{1986}]{1986ApJ...305L..61D}
{Djorgovski} S.,  {King} I.~R.,  1986, \apjl, 305, L61

\bibitem[\protect\citeauthoryear{{Fall} \& {Rees}}{{Fall} \&
  {Rees}}{1977}]{1977MNRAS.181P..37F}
{Fall} S.~M.,  {Rees} M.~J.,  1977, \mnras, 181, 37P

\bibitem[\protect\citeauthoryear{{Fall} \& {Zhang}}{{Fall} \&
  {Zhang}}{2001}]{2001ApJ...561..751F}
{Fall} S.~M.,  {Zhang} Q.,  2001, \apj, 561, 751

\bibitem[\protect\citeauthoryear{{Fukushige} \& {Heggie}}{{Fukushige} \&
  {Heggie}}{2000}]{2000MNRAS.318..753F}
{Fukushige} T.,  {Heggie} D.~C.,  2000, \mnras, 318, 753

\bibitem[\protect\citeauthoryear{{Georgiev}, {Puzia}, {Hilker} \&
  {Goudfrooij}}{{Georgiev} et~al.}{2009}]{2009MNRAS.392..879G}
{Georgiev} I.~Y.,  {Puzia} T.~H.,  {Hilker} M.,    {Goudfrooij} P.,  2009,
  \mnras, 392, 879

\bibitem[\protect\citeauthoryear{{Gieles} \& {Baumgardt}}{{Gieles} \&
  {Baumgardt}}{2008}]{2008MNRAS.389L..28G}
{Gieles} M.,  {Baumgardt} H.,  2008, \mnras, 389, L28

\bibitem[\protect\citeauthoryear{{Gieles}, {Baumgardt}, {Heggie} \&
  {Lamers}}{{Gieles} et~al.}{2010}]{2010MNRAS.408L..16G}
{Gieles} M.,  {Baumgardt} H.,  {Heggie} D.~C.,    {Lamers} H.~J.~G.~L.~M.,
  2010, \mnras, 408, L16

\bibitem[\protect\citeauthoryear{{Giersz} \& {Heggie}}{{Giersz} \&
  {Heggie}}{1994}]{1994MNRAS.268..257G}
{Giersz} M.,  {Heggie} D.~C.,  1994, \mnras, 268, 257

\bibitem[\protect\citeauthoryear{{Giersz} \& {Heggie}}{{Giersz} \&
  {Heggie}}{1996}]{1996MNRAS.279.1037G}
{Giersz} M.,  {Heggie} D.~C.,  1996, \mnras, 279, 1037

\bibitem[\protect\citeauthoryear{{Giersz} \& {Heggie}}{{Giersz} \&
  {Heggie}}{2011}]{2011MNRAS.410.2698G}
{Giersz} M.,  {Heggie} D.~C.,  2011, \mnras, 410, 2698

\bibitem[\protect\citeauthoryear{{Gnedin} \& {Ostriker}}{{Gnedin} \&
  {Ostriker}}{1997}]{1997ApJ...474..223G}
{Gnedin} O.~Y.,  {Ostriker} J.~P.,  1997, \apj, 474, 223

\bibitem[\protect\citeauthoryear{{Goodman}}{{Goodman}}{1984}]{1984ApJ...280..2%
98G}
{Goodman} J.,  1984, \apj, 280, 298

\bibitem[\protect\citeauthoryear{{G{\"u}rkan}, {Freitag} \&
  {Rasio}}{{G{\"u}rkan} et~al.}{2004}]{2004ApJ...604..632G}
{G{\"u}rkan} M.~A.,  {Freitag} M.,    {Rasio} F.~A.,  2004, \apj, 604, 632

\bibitem[\protect\citeauthoryear{{Harris}}{{Harris}}{1991}]{1991ARA&A..29..543%
H}
{Harris} W.~E.,  1991, \araa, 29, 543

\bibitem[\protect\citeauthoryear{{Harris}}{{Harris}}{1996}]{1996AJ....112.1487%
H}
{Harris} W.~E.,  1996, \aj, 112, 1487

\bibitem[\protect\citeauthoryear{{Harris}, {Spitler}, {Forbes} \&
  {Bailin}}{{Harris} et~al.}{2010}]{2010MNRAS.401.1965H}
{Harris} W.~E.,  {Spitler} L.~R.,  {Forbes} D.~A.,    {Bailin} J.,  2010,
  \mnras, 401, 1965

\bibitem[\protect\citeauthoryear{{Heggie} \& {Giersz}}{{Heggie} \&
  {Giersz}}{2008}]{2008MNRAS.389.1858H}
{Heggie} D.~C.,  {Giersz} M.,  2008, \mnras, 389, 1858

\bibitem[\protect\citeauthoryear{{Heggie}, {Giersz}, {Spurzem} \&
  {Takahashi}}{{Heggie} et~al.}{1998}]{1998HiA....11..591H}
{Heggie} D.~C.,  {Giersz} M.,  {Spurzem} R.,    {Takahashi} K.,  1998,
  Highlights of Astronomy, 11, 591

\bibitem[\protect\citeauthoryear{{H{\'e}non}}{{H{\'e}non}}{1961}]{1961AnAp...2%
4..369H}
{H{\'e}non} M.,  1961, Annales d'Astrophysique, 24, 369

\bibitem[\protect\citeauthoryear{{H{\'e}non}}{{H{\'e}non}}{1965}]{1965AnAp...2%
8...62H}
{H{\'e}non} M.,  1965, Annales d'Astrophysique, 28, 62

\bibitem[\protect\citeauthoryear{{H{\'e}non}}{{H{\'e}non}}{1975}]{1975IAUS...6%
9..133H}
{H{\'e}non} M.,  1975, in {A.~Hayli,} ed., Proc. IAU Symp. 69, Dynamics of the
  Solar Systems. Reidel, Dordrecht, p. 133 {}

\bibitem[\protect\citeauthoryear{{Hills}}{{Hills}}{1980}]{1980ApJ...235..986H}
{Hills} J.~G.,  1980, \apj, 235, 986

\bibitem[\protect\citeauthoryear{{Hurley}}{{Hurley}}{2007}]{2007MNRAS.379...93%
H}
{Hurley} J.~R.,  2007, \mnras, 379, 93

\bibitem[\protect\citeauthoryear{{Hurley} \& {Mackey}}{{Hurley} \&
  {Mackey}}{2010}]{2010MNRAS.408.2353H}
{Hurley} J.~R.,  {Mackey} A.~D.,  2010, \mnras, 408, 2353

\bibitem[\protect\citeauthoryear{{Illingworth} \& {King}}{{Illingworth} \&
  {King}}{1977}]{1977ApJ...218L.109I}
{Illingworth} G.,  {King} I.~R.,  1977, \apjl, 218, L109

\bibitem[\protect\citeauthoryear{{Inagaki} \& {Lynden-Bell}}{{Inagaki} \&
  {Lynden-Bell}}{1983}]{1983MNRAS.205..913I}
{Inagaki} S.,  {Lynden-Bell} D.,  1983, \mnras, 205, 913

\bibitem[\protect\citeauthoryear{{Jord{\'a}n}}{{Jord{\'a}n}}{2004}]{2004ApJ...%
613L.117J}
{Jord{\'a}n} A.,  2004, \apjl, 613, L117

\bibitem[\protect\citeauthoryear{{Jord{\'a}n}, {C{\^o}t{\'e}}, {Blakeslee},
  {Ferrarese}, {McLaughlin}, {Mei}, {Peng}, {Tonry}, {Merritt},
  {Milosavljevi{\'c}}, {Sarazin}, {Sivakoff} \& {West}}{{Jord{\'a}n}
  et~al.}{2005}]{2005ApJ...634.1002J}
{Jord{\'a}n} A.,  {C{\^o}t{\'e}} P.,  {Blakeslee} J.~P.,  {Ferrarese} L.,
  {McLaughlin} D.~E.,  {Mei} S.,  {Peng} E.~W.,  {Tonry} J.~L.,  {Merritt} D.,
  {Milosavljevi{\'c}} M.,  {Sarazin} C.~L.,  {Sivakoff} G.~R.,    {West} M.~J.,
   2005, \apj, 634, 1002

\bibitem[\protect\citeauthoryear{{Jord{\'a}n}, {McLaughlin}, {C{\^o}t{\'e}},
  {Ferrarese}, {Peng}, {Mei}, {Villegas}, {Merritt}, {Tonry} \&
  {West}}{{Jord{\'a}n} et~al.}{2007}]{2007ApJS..171..101J}
{Jord{\'a}n} A.,  {McLaughlin} D.~E.,  {C{\^o}t{\'e}} P.,  {Ferrarese} L.,
  {Peng} E.~W.,  {Mei} S.,  {Villegas} D.,  {Merritt} D.,  {Tonry} J.~L.,
  {West} M.~J.,  2007, \apjs, 171, 101

\bibitem[\protect\citeauthoryear{{Kim}, {Lee} \& {Goodman}}{{Kim}
  et~al.}{1998}]{1998ApJ...495..786K}
{Kim} S.~S.,  {Lee} H.~M.,    {Goodman} J.,  1998, \apj, 495, 786

\bibitem[\protect\citeauthoryear{{Kravtsov} \& {Gnedin}}{{Kravtsov} \&
  {Gnedin}}{2005}]{2005ApJ...623..650K}
{Kravtsov} A.~V.,  {Gnedin} O.~Y.,  2005, \apj, 623, 650

\bibitem[\protect\citeauthoryear{{K{\"u}pper}, {Kroupa} \&
  {Baumgardt}}{{K{\"u}pper} et~al.}{2008}]{2008MNRAS.389..889K}
{K{\"u}pper} A.~H.~W.,  {Kroupa} P.,    {Baumgardt} H.,  2008, \mnras, 389, 889

\bibitem[\protect\citeauthoryear{{K{\"u}pper}, {Kroupa}, {Baumgardt} \&
  {Heggie}}{{K{\"u}pper} et~al.}{2010}]{2010MNRAS.407.2241K}
{K{\"u}pper} A.~H.~W.,  {Kroupa} P.,  {Baumgardt} H.,    {Heggie} D.~C.,  2010,
  \mnras, 407, 2241

\bibitem[\protect\citeauthoryear{{Lamers}, {Baumgardt} \& {Gieles}}{{Lamers}
  et~al.}{2010}]{2010MNRAS.409..305L}
{Lamers} H.~J.~G.~L.~M.,  {Baumgardt} H.,    {Gieles} M.,  2010, \mnras, 409,
  305

\bibitem[\protect\citeauthoryear{{Lee}}{{Lee}}{2002}]{2002IAUS..207..584L}
{Lee} H.~M.,  2002, in {D.~P.~Geisler, E.~K.~Grebel, \& D.~Minniti} ed., Proc.
  IAU Symp. 207, Extragalactic Star Clusters, Astron. Soc. Pac., San Francisco,
  p. 584 {}

\bibitem[\protect\citeauthoryear{{Lee}, {Fahlman} \& {Richer}}{{Lee}
  et~al.}{1991}]{1991ApJ...366..455L}
{Lee} H.~M.,  {Fahlman} G.~G.,    {Richer} H.~B.,  1991, \apj, 366, 455

\bibitem[\protect\citeauthoryear{{Lee} \& {Goodman}}{{Lee} \&
  {Goodman}}{1995}]{1995ApJ...443..109L}
{Lee} H.~M.,  {Goodman} J.,  1995, \apj, 443, 109

\bibitem[\protect\citeauthoryear{{Lee} \& {Ostriker}}{{Lee} \&
  {Ostriker}}{1987}]{1987ApJ...322..123L}
{Lee} H.~M.,  {Ostriker} J.~P.,  1987, \apj, 322, 123

\bibitem[\protect\citeauthoryear{{Lynden-Bell} \& {Wood}}{{Lynden-Bell} \&
  {Wood}}{1968}]{1968MNRAS.138..495L}
{Lynden-Bell} D.,  {Wood} R.,  1968, \mnras, 138, 495

\bibitem[\protect\citeauthoryear{{Mackey}, {Huxor}, {Ferguson}, {Irwin},
  {Tanvir}, {McConnachie}, {Ibata}, {Chapman} \& {Lewis}}{{Mackey}
  et~al.}{2010}]{2010ApJ...717L..11M}
{Mackey} A.~D.,  {Huxor} A.~P.,  {Ferguson} A.~M.~N.,  {Irwin} M.~J.,  {Tanvir}
  N.~R.,  {McConnachie} A.~W.,  {Ibata} R.~A.,  {Chapman} S.~C.,    {Lewis}
  G.~F.,  2010, \apjl, 717, L11

\bibitem[\protect\citeauthoryear{{McLaughlin}}{{McLaughlin}}{2000}]{2000ApJ...%
539..618M}
{McLaughlin} D.~E.,  2000, \apj, 539, 618

\bibitem[\protect\citeauthoryear{{McLaughlin}, {Barmby}, {Harris}, {Forbes} \&
  {Harris}}{{McLaughlin} et~al.}{2008}]{2008MNRAS.384..563M}
{McLaughlin} D.~E.,  {Barmby} P.,  {Harris} W.~E.,  {Forbes} D.~A.,    {Harris}
  G.~L.~H.,  2008, \mnras, 384, 563

\bibitem[\protect\citeauthoryear{{McLaughlin} \& {Fall}}{{McLaughlin} \&
  {Fall}}{2008}]{2008ApJ...679.1272M}
{McLaughlin} D.~E.,  {Fall} S.~M.,  2008, \apj, 679, 1272

\bibitem[\protect\citeauthoryear{{McLaughlin} \& {van der Marel}}{{McLaughlin}
  \& {van der Marel}}{2005}]{2005ApJS..161..304M}
{McLaughlin} D.~E.,  {van der Marel} R.~P.,  2005, \apjs, 161, 304

\bibitem[\protect\citeauthoryear{{Niederste-Ostholt}, {Belokurov}, {Evans},
  {Koposov}, {Gieles} \& {Irwin}}{{Niederste-Ostholt}
  et~al.}{2010}]{2010MNRAS.408L..66N}
{Niederste-Ostholt} M.,  {Belokurov} V.,  {Evans} N.~W.,  {Koposov} S.,
  {Gieles} M.,    {Irwin} M.~J.,  2010, \mnras, 408, L66

\bibitem[\protect\citeauthoryear{{Ostriker} \& {Gnedin}}{{Ostriker} \&
  {Gnedin}}{1997}]{1997ApJ...487..667O}
{Ostriker} J.~P.,  {Gnedin} O.~Y.,  1997, \apj, 487, 667

\bibitem[\protect\citeauthoryear{{Pe{\~n}arrubia}, {Walker} \&
  {Gilmore}}{{Pe{\~n}arrubia} et~al.}{2009}]{2009MNRAS.399.1275P}
{Pe{\~n}arrubia} J.,  {Walker} M.~G.,    {Gilmore} G.,  2009, \mnras, 399, 1275

\bibitem[\protect\citeauthoryear{{Portegies Zwart}, {Gaburov}, {Chen} \&
  {G{\"u}rkan}}{{Portegies Zwart} et~al.}{2007}]{2007MNRAS.378L..29P}
{Portegies Zwart} S.,  {Gaburov} E.,  {Chen} H.,    {G{\"u}rkan} M.~A.,  2007,
  \mnras, 378, L29

\bibitem[\protect\citeauthoryear{{Prieto} \& {Gnedin}}{{Prieto} \&
  {Gnedin}}{2008}]{2008ApJ...689..919P}
{Prieto} J.~L.,  {Gnedin} O.~Y.,  2008, \apj, 689, 919

\bibitem[\protect\citeauthoryear{{Spitzer}}{{Spitzer}}{1987}]{1987degc.book...%
..S}
{Spitzer} L.,  1987, {Dynamical evolution of globular clusters}.
Princeton, NJ, Princeton University Press, 1987, 191 p.

\bibitem[\protect\citeauthoryear{{Spitzer}}{{Spitzer}}{1969}]{1969ApJ...158L.1%
39S}
{Spitzer} L.~J.,  1969, \apjl, 158, L139

\bibitem[\protect\citeauthoryear{{van den Bergh} \& {Mackey}}{{van den Bergh}
  \& {Mackey}}{2004}]{2004MNRAS.354..713V}
{van den Bergh} S.,  {Mackey} A.~D.,  2004, \mnras, 354, 713

\bibitem[\protect\citeauthoryear{{van den Bergh}, {Morbey} \& {Pazder}}{{van
  den Bergh} et~al.}{1991}]{1991ApJ...375..594V}
{van den Bergh} S.,  {Morbey} C.,    {Pazder} J.,  1991, \apj, 375, 594

\bibitem[\protect\citeauthoryear{{Vesperini} \& {Heggie}}{{Vesperini} \&
  {Heggie}}{1997}]{1997MNRAS.289..898V}
{Vesperini} E.,  {Heggie} D.~C.,  1997, \mnras, 289, 898

\bibitem[\protect\citeauthoryear{{Villegas}, {Jord{\'a}n}, {Peng}, {Blakeslee},
  {C{\^o}t{\'e}}, {Ferrarese}, {Kissler-Patig}, {Mei}, {Infante}, {Tonry} \&
  {West}}{{Villegas} et~al.}{2010}]{2010ApJ...717..603V}
{Villegas} D.,  {Jord{\'a}n} A.,  {Peng} E.~W.,  {Blakeslee} J.~P.,
  {C{\^o}t{\'e}} P.,  {Ferrarese} L.,  {Kissler-Patig} M.,  {Mei} S.,
  {Infante} L.,  {Tonry} J.~L.,    {West} M.~J.,  2010, \apj, 717, 603

\bibitem[\protect\citeauthoryear{{Williams}, {Ciardullo}, {Durrell},
  {Feldmeier}, {Sigurdsson}, {Vinciguerra}, {Jacoby}, {von Hippel}, {Ferguson},
  {Tanvir}, {Arnaboldi}, {Gerhard}, {Aguerri} \& {Freeman}}{{Williams}
  et~al.}{2007}]{2007ApJ...654..835W}
{Williams} B.~F.,  {Ciardullo} R.,  {Durrell} P.~R.,  {Feldmeier} J.~J.,
  {Sigurdsson} S.,  {Vinciguerra} M.,  {Jacoby} G.~H.,  {von Hippel} T.,
  {Ferguson} H.~C.,  {Tanvir} N.~R.,  {Arnaboldi} M.,  {Gerhard} O.,  {Aguerri}
  J.~A.~L.,    {Freeman} K.~C.,  2007, \apj, 654, 835

\bibitem[\protect\citeauthoryear{{Zonoozi}, {K{\"u}pper}, {Baumgardt}, {Haghi},
  {Kroupa} \& {Hilker}}{{Zonoozi} et~al.}{2011}]{2011MNRAS.411.1989Z}
{Zonoozi} A.~H.,  {K{\"u}pper} A.~H.~W.,  {Baumgardt} H.,  {Haghi} H.,
  {Kroupa} P.,    {Hilker} M.,  2011, \mnras, 411, 1989

\end{thebibliography}

%\onecolumn
\appendix
%%%%%%%%%%%%%%%%%%%%%%%%%%%%%%%%%%%%%%%%%%%%
\section{Alternative escape criterion}

\label{A}
In Section~\ref{sec:cycle} we have solved the simple case in which the
escape rate is constant during the entire evolution.  This resulted
from the fact that we used the scaling $\xi\propto\tcr$, cancelling
out the $\tcr$ term of $\trh$ in equation~(\ref{eq:xi}).  The constant
escape rate is to first order a good approximation, but a slightly
improved description for $\xi$ would be the exponential function
mentioned in Section~\ref{ssec:relative}.  In principle all
derivations of Section~\ref{sec:cycle} can be repeated using different
functional forms for $\xi$.  Together with the constraint that
$\zeta=\,$constant we can always solve for $\chi$. For the exponential
function mentioned before we would have to proceed with numerical
integrations of the various differential equations, which we will not
do here.

Stars that have enough energy to escape take a finite time to do so,
and this has a profound effect on the escape rate.  This time scale
can be very long for stars with energies only slightly higher than the
escape energy \citep{2000MNRAS.318..753F}. We will not discuss the
details of the theory here and instead proceed directly to the
consequences for the escape rate. This was considered by
\citet{2001MNRAS.325.1323B} and he has shown that the relevant
time-scale for escape is in fact a combination of $\trh$ and $\tcr$
\begin{eqnarray}
\tml&\propto&\trh^x\tcr^{1-x}\\
      &\propto&\trh\left(\frac{\tcr}{\trh}\right)^{1-x},
\end{eqnarray}
 i.e.
\begin{equation}
\tml \simeq
      \trh\left(\frac{N}{\nref}\right)^{x-1},
      \label{eq:tesc} 
\end{equation}
with $x\simeq3/4$. The proportionality, i.e. the value of the
reference number $\nref$, may be determined by matching the time
scales following from the theory to the results of $N$-body
simulations.  We again define the dimensionless escape-rate $\xi$ as
the fraction of stars lost per relaxation time such that an additional
(small) $N$-dependence needs to be included; thus instead of
equation~(\ref{eq:xi}) we have

\begin{equation}
\xi\equiv-\frac{\ndot\trh}{N}=\frac{3}{5}\zeta\left(\frac{\tcr}{\tcrmax}\right)\left(\frac{N}{\nref}\right)^{1-x}.
\label{eq:xi_b01}
\end{equation}
The escape rate corrected for the tidal density then becomes
\begin{equation}
\ndot\tcrj=-\frac{71\zeta}{\rhrjmaxtext^{3/2}}\left(\frac{N}{\nref}\right)^{1-x}
\label{eq:ndot2}
\end{equation}
in place of equation~(\ref{eq:tcrjndotn}). Consequently the
$N$-dependence also enters in the expression for $\chi$ through the
constraint that $\zeta$ is constant (equation~\ref{eq:zeta})
\begin{equation}
\chi\equiv\frac{\tcrdot\trh}{\tcr}=\frac{3}{2}\zeta\left(1-\frac{\tcr}{\tcrmax}\left[\frac{N}{\nref}\right]^{1-x}\right),
\label{eq:chi_b01}
\end{equation}
in place of equation~(\ref{eq:chi}).

%________________________________________________
\subsection{Motion in the $\tcr-N$ plane}
\label{subsec:tcr_n2}
The motion in the $\tcr-N$ plane can be found in a similar way as in
Section~\ref{subsec:tcr_n}. With $\ndot$ and $\tcrdot$ defined as
 in equations~(\ref{eq:xi_b01}) and
(\ref{eq:chi_b01}) we then find

\begin{eqnarray}
\frac{\dr N}{\dr  \tcr}&=&-\frac{\xi}{\chi}\frac{N}{\tcr}\nonumber\\
                       &=&\frac{2}{5}\frac{N^{2-x}}{\tcr  N^{1-x}-\tcrmax\nref^{1-x}}.
\end{eqnarray}
 Because the variables $\tcr$ and $N$ can not be separated, a variable
 substitution $u=\tcr N^{1-x}$ needs to be performed and then the
 variables $u$ and $N$ can be separated and the integration can be
 performed. The result is
\begin{equation}
\tcr=\frac{\tcrmax}{A}\left(\frac{N}{\nref}\right)^{x-1}\left(1-\left[\frac{N}{N_0}\right]^{5A/2}\right).
\label{eq:tcr_n_b01}
\end{equation}
Here $A\equiv1+(2/5)(1-x)$ and we again used $\tcrinit=0$.  When
$x=1$, then $A=1$ and equation~(\ref{eq:tcr_n_b01}) reduces to the
result of Section~\ref{sec:cycle} (equation~\ref{eq:tcr_n}).

The continuous growth of $\tcr$ with decreasing $N$ in the tidal
regime shows that clusters are not evolving homologically any more in
the evaporation dominated regime. This is different from the model of
\henon\ where $\tcr$ becomes a constant related to the tidal density.
The time-dependence also changes and this is discussed in
\ref{ssec:time-dependence2}.

%_______________________________________
\subsection{Time-dependence}
\label{ssec:time-dependence2}
The time-dependence of the evolution can be found as before from an
integration over $\ndot$ (equation~\ref{eq:ndot2})
\begin{equation}
N(t)=N_0\left(1-\frac{t}{\tevn}\right)^{1/x},
\end{equation}
with $\tevn$ now defined as
\begin{equation}
\tevn=\frac{\rhrjmaxtext^{3/2}}{71 \zeta }\frac{\nref^{{1-x}}}{x}\tcrj N_0^x.
\label{eq:tev2}
\end{equation}
The time-dependent $\tcr(t)$ follows from
equation~(\ref{eq:tcr_n_b01}) and the above.  We can then also get
$\rh(t)$ and $\trh(t)$, but this will not be done here.

%%%%%%%%%%%%%%%%%%%%%%%%%%%%%%%%%%%%%%%%%%%%
\section{Tracks and isochrones in units of $M$, $\rhoh$ and $\rg$}
\label{B}

In this appendix we present the results of Section~\ref{sec:cycle} in
terms of $M$, $\rhoh$ and $\rhoj$.  We return to the case $x = 1$.

%_____________________________________________________________________________________________
\subsection{Tracks}
The evolution of mass is 
\begin{equation}
M(M_0,t)=M_0-\mdotabs t
\end{equation}
If we use $\mmean=0.5$ then from equations (\ref{eq:tcrmax}) and~(\ref{eq:ndottcrj}) we find
\begin{equation}
\mdotabs\simeq35.4\zeta\left(\frac{
  G\rhoj}{\rhrjmaxtexts^{3}}\right)^{1/2}.
\label{eq:mdot-appendix}
\end{equation}
Together with equations (\ref{eq:nt}) and~(\ref{eq:tcrt}) we then have
for the density

\begin{eqnarray}
\rhoh(\rhoj,M_0,t)\hsp&=&\hsp\frac{1}{2}\rhrjmax^{-3}\rhoj\left(1-\left[1-\frac{\mdotabs t}{M_0}\right]^{5/2}\right)^{-2}\!\!\!,\\
\rhoh(M_0,t)\hsp&\simeq&\hsp\frac{1}{G}\left(\frac{M_0}{125\zeta t}\right)^2,\,\,\,M_0\gg\mdotabs t,\\ 
\rhoh(\rhoj,M_0,t)\hsp&\simeq&\hsp\frac{1}{2}\rhrjmax^{-3}\rhoj,\,\,\,M_0\simeq\mdotabs t,
\end{eqnarray}
or in terms of radius

\begin{eqnarray}
\rh(\rhoj,M_0,t)\hsp&=&\hsp\left(\frac{6}{8\pi}\right)^{1/3}\rhrjmax\left(\frac{M_0}{\rhoj}\right)^{1/3}\left(1-\frac{\mdotabs t}{M_0}\right)^{1/3}\nonumber\\
& & \,\,\,\,\times\left(1-\left[1-\frac{\mdotabs t}{M_0}\right]^{5/2}\right)^{2/3},\\
\rh(M_0,t)\hsp&\simeq&\hsp\left(\frac{3G}{8\pi M_0}\right)^{1/3}\left(125\zeta t\right)^{2/3},\,\,\,M_0\gg\mdotabs t,\\ 
\rh(\rhoj,M_0,t)\hsp&\simeq&\hsp\left(\frac{6}{8\pi}\right)^{1/3}\rhrjmax\left(\frac{M_0}{\rhoj}\right)^{1/3}\nonumber\\
& &\times\left(1-\frac{\mdot t}{M_0}\right)^{1/3},\,\,\,M_0\simeq\mdotabs t.
\end{eqnarray}

%_____________________________________________________________________________________________
\subsection{Isochrones}
The isochrones are easily found from the tracks and using $M_0=M+\mdotabs t$
\begin{eqnarray}
\rhoh(\rhoj,M,t)\hsp&=&\hsp\frac{1}{2}\rhrjmax^{-3}\rhoj\left(1-\left[1+\frac{\mdotabs t}{M}\right]^{-5/2}\right)^{-2}\hsp,\\
\rhoh(M,t)\hsp&\simeq&\hsp\frac{1}{G}\left(\frac{M}{125\zeta t}\right)^2,\,\,\,M\gg\mdotabs t,\\
\rhoh(\rhoj,M)\hsp&\simeq&\hsp\frac{1}{2}\rhrjmax^{-3}\rhoj,\,\,\,M\simeq\mdotabs t,
\end{eqnarray}
or in terms of radius
\begin{eqnarray}
\rh(\rhoj,M,t)&=&\left(\frac{6}{8\pi}\right)^{1/3}\rhrjmax\left(\frac{M}{\rhoj}\right)^{1/3}\nonumber\\
& &\times\left(1-\left[1+\frac{\mdotabs t}{M}\right]^{-5/2}\right)^{2/3},\\
\rh(M,t)	\hsp&\simeq&\hsp\left(\frac{3G}{8\pi M}\right)^{1/3}\left(125\zeta t\right)^{2/3},M\gg\mdotabs t,\\ 
\rh(\rhoj,M) \hsp&\simeq&\hsp\left(\frac{6}{8\pi}\right)^{1/3}\rhrjmax\left(\frac{M}{\rhoj}\right)^{1/3}\hsp, M\simeq\mdotabs t.
\end{eqnarray}

%_____________________________________________________________________________________________
\subsection{In the isothermal halo approximation}
In an isothermal halo with constant circular velocity $\vg$ the
density within the Jacobi radius depends on $\rg$ as
\begin{eqnarray}
\rhoj&=&\frac{3}{2\pi G}\frac{\vg^2}{\rg^2},\\
        &\simeq&5.376\left(\frac{\rg}{\kpc}\right)^{-2}.
\end{eqnarray}
In the last step we used $\vg=220\,\kms$.  If we use the constants as
described in Section~\ref{sec:cycle}: $\rhrjmaxtext=0.145$,
$\zeta=0.2$ and
$G\simeq4.5\times10^{-3}\,\pc^3\,\msun^{-1}\,\myr^{-2}$ then from
equation~(\ref{eq:mdot-appendix}) we obtain
\begin{equation}
\mdotabs\simeq20\,\msun\,\myr^{-1}\left(\frac{ \rg}{\kpc}\right)^{-1}.
\end{equation}
The tracks and isochrones for $\rhoh$ are shown in
Section~\ref{sec:obs} using this isothermal halo approximation. For
the isochrones an age of $13\,\gyr$ was used.

%%%%%%%%%%%%%%%%%%%%%%%%%%%%%%%%%%%%%%%%%%%%
\section{Tracks and isochrones in units of $M$, $\rhoh$ and $\rg$, including the escape time effect}
\label{C}

If we include the effect of the escape time (Appendix~\ref{A}) we get
for the mass evolution
\begin{equation}
M(M_0,t)=\left(M_0^x-\mdotx t\right)^{1/x},
\label{eq:mtx}
\end{equation} 
where
\begin{equation}
\mdotx\simeq35.4\zeta\left(\frac{ G\rhoj}{\rhrjmaxtexts^{3}}\right)^{1/2}\frac{x}{\mref^{1-x}}.  
\end{equation}
Here $\mref\equiv\mmean\nref$ is a constant reference mass that has a
similar role as $\nref$ introduced in Appendix~\ref{A}, and we again
adopt $\mmean = 0.5\msun$. Note that $\mdotx$ is not anymore the
time-derivative of $M$.
  
%____________________________________________________________________
\subsection{Tracks}
 Using equation~(\ref{eq:tcrmax}) for $\tcrmax$, we find that the
tracks for  general $x$ are 
\begin{eqnarray}
\rhoh(\rhoj,M_0,t)\hsp&=&\hsp\frac{A^2}{2}\rhrjmax^{-3}\hsp\rhoj\left(\frac{M_0}{\mref}\right)^{2-2x}\hsp\left(1-\frac{\mdotx t}{M_0^x}\right)^{2/x-2}\nonumber\\
& &\times\left(1-\left[1-\frac{\mdotx t}{M_0^x}\right]^{(7/2)/x-1}\right)^{-2},\\
\rhoh(M_0,t)&\simeq&\frac{1}{G}\left(\frac{M_0}{125\zeta t}\right)^2,\,\,\,M_0^x\gg\mdotx t,\\ 
\rhoh(\rhoj,M_0,t)&\simeq&\frac{A^2}{2}\rhrjmax^{-3}\rhoj\left(\frac{M_0}{\mref}\right)^{2-2x}\nonumber\\
& &\times\left(1-\frac{\mdotx t}{M_0^x}\right)^{2/x-2},\,\,\,M_0^x\simeq\mdotx t,
\end{eqnarray}
or in terms of radius
\begin{eqnarray}
\rh(\rhoj,M_0,t)\hsp&=&\hsp\left(\frac{6}{8\pi A^2}\right)^{1/3}\rhrjmax\left(\frac{M_0}{\rhoj}\right)^{1/3}\left(\frac{M_0}{\mref}\right)^{(2x-2)/3}\nonumber\\
 \times&&\hsp\hsp\hsp\hsp\left(1-\frac{\mdotx t}{M_0^x}\right)^{\!\!(2-1/x)/3}\!\!\left(1-\left[1-\frac{\mdotx t}{M_0^x}\right]^{(7/2)/x-1}\right)^{\!\!2/3}\hsp\hsp,\nonumber\\
& &\\
\rh(M_0,t)\hsp&\simeq&\hsp\left(\frac{3G}{8\pi M_0}\right)^{1/3}\left(125\zeta t\right)^{2/3},\,\,\,M_0^x\gg\mdotx t,\\ 
\rh(\rhoj,M_0,t)\hsp&\simeq&\hsp\left(\frac{6}{8\pi A^2}\right)^{1/3}\rhrjmax\left(\frac{M_0}{\rhoj}\right)^{1/3}\left(\frac{M_0}{\mref}\right)^{(2x-2)/3}\nonumber\\
 & &\times\left(1-\frac{\mdotx t}{M_0^x}\right)^{(2-1/x)/3},\,\,\,M_0^x\simeq\mdotx t.
\end{eqnarray}
%_____________________________________________________________________________________________
\subsection{Isochrones}
The isochrones for variable $x$ can be found from the tracks by
insertion of an expression for $M_0(M,\mdotx,t)$
(equation~\ref{eq:mtx})
\begin{eqnarray}
\rhoh(\rhoj,M,t)\hsp&=&\hsp\frac{A^2}{2}\rhrjmax^{-3}\rhoj\left(\frac{M}{\mref}\right)^{2-2x}\nonumber\\
& &\times\left(1-\left[1+\frac{\mdotx t}{M^x}\right]^{1-(7/2)/x}\right)^{-2},\\
\rhoh(M,t)	\hsp&\simeq&\hsp\frac{1}{G}\left(\frac{M}{125\zeta t}\right)^2,\,\,\,M^x\gg\mdotx t,\\
\rhoh(\rhoj,M)\hsp&\simeq&\hsp\frac{A^2}{2}\rhrjmax^{-3}\hsp\rhoj\left(\frac{M}{\mref}\right)^{2-2x}\hsp,\,\,M^x\simeq\mdotx t,
\end{eqnarray}
or in terms of radius
\begin{eqnarray}
\rh(\rhoj,M,t)&=&\left(\frac{6}{8\pi A^2}\right)^{1/3}\rhrjmax\left(\frac{M}{\rhoj}\right)^{1/3}\left(\frac{M}{\mref}\right)^{(2x-2)/3}\nonumber\\
& &\times\left(1-\left[1+\frac{\mdotx t}{M^x}\right]^{1-(7/2)/x}\right)^{2/3},\\
\rh(M,t)	&\simeq&\left(\frac{3G}{8\pi M}\right)^{1/3}\left(125\zeta t\right)^{2/3},\,\,\,M^x\gg\mdotx t,\\ 
\rh(\rhoj,M)&\simeq&\left(\frac{6}{8\pi A^2}\right)^{1/3}\rhrjmax\left(\frac{M}{\rhoj}\right)^{1/3}\nonumber\\
& &\times\left(\frac{M}{\mref}\right)^{(2x-2)/3}, \,\,\,M^x\simeq\mdotx t.
\end{eqnarray}
%_______________________________________
\subsection{Implications for the comparison to Milky Way globular clusters}
The main change that follows from the introduction of $x\ne1$ occurs
in the evaporation dominated phase where we do not have a constant
$\rh/\rj$ anymore, but instead

\begin{equation}
\rhoh/\rhoj\propto M^{2-2x},
\end{equation}
or 
\begin{equation}
\rh/\rj\propto M^{2(x-1)/3}.
\end{equation}
 For $x=3/4$ and the
isothermal halo approximation these scaling relations are equivalent
to 

\begin{eqnarray}
\rhoh\propto M^{1/2}\rg^{-2},
\end{eqnarray}
or
\begin{eqnarray}
\rh\propto M^{1/6}\rg^{2/3}.
\end{eqnarray}

So at a given $\rg$ this relation implies that $\rh\propto M^{1/6}$,
depending on the exact value of $x$\footnote{A relation $\rh\propto
  M^0$ is found for $x=1/2$.}. If we would have taken the Coulomb
logarithm into account, the index of $1/6$ would be slightly smaller.
So the particular scaling of $\tml\propto\trh^{3/4}$ has as a result
that in the regime where mass loss dominates the cluster half-mass
radius is (nearly) independent of $M$ and is determined mainly by the
tidal field strength.  In the expansion dominated phase we still find
$\tcr\propto N^{-1}$, or $\rhoh\propto M^2$, at a given age.

\end{document}